\documentclass[11pt]{article}
\usepackage{graphicx}

\begin{document}
\oddsidemargin .03in
\evensidemargin 0 true pt
\topmargin -.4in

\def\ra{{\rightarrow}}
\def\a{{\alpha}}
\def\b{{\beta}}
\def\l{{\lambda}}
\def\eps{{\epsilon}}
\def\pr{{\partial}}
\def\tri{{\triangle}}
\def\na{{\nabla }}
\def\sp{\vspace{.15in}}
\def\hs{\hspace{.25in}}
\def\n{\nonumber}
\def\ni{{\noindent}}
\def\Ra{{\Rightarrow}}

\newcommand{\be}{\begin{equation}} \newcommand{\ee}{\end{equation}}
\newcommand{\bea}{\begin{eqnarray}}\newcommand{\eea}{\end{eqnarray}}


\topmargin= -.2in
\textheight 9.5in

\begin{flushright}
{\bf arXiv: 1407.7756 [hep-th]}
\end{flushright}

\baselineskip= 18 truept

\vspace{.3in}

\centerline{\Large\bf Quantum effects in topological and Schwarzschild de Sitter brane:}
\centerline{\Large\bf Aspects of torsion on ${\mathbf{(D{\bar D})_4}}$-brane universe}

\vspace{.6in}
\noindent
\centerline{\bf Richa Kapoor, Supriya Kar and Deobrat Singh}

\vspace{.13in}

\noindent
\centerline{{\Large Department of Physics \& Astrophysics}}
\centerline{\Large University of Delhi, New Delhi 110 007, India}

\vspace{.2in}
\begin{center}
{\today}
\end{center}

\thispagestyle{empty}

\vspace{.6in}

\begin{abstract}
We investigate an effective torsion curvature in a second order formalism underlying a two form world-volume dynamics in a $D_5$-brane. In particular, we consider the two form in presence of a background (open string) metric in a $U(1)$ gauge theory. Interestingly the formalism may be viewed via a 
non-coincident pair of $(D{\bar D})_5$-brane with a global NS two form on an anti brane and a local two form on a brane. The energy-momentum tensor is computed in the six dimensional CFT. It is shown to source a metric fluctuation on a vacuum created pair of $(D{\bar D})_4$-brane at a cosmological horizon by the two form quanta in the gauge theory. The emergent gravity scenario is shown to describe a low energy (perturbative) string vacuum in $6D$ with a (non-perturbative) quantum correction by a lower ($p<5$) dimensional $D_p$ brane or an anti brane in the formalism. A closed string exchange between a pair of $(D{\bar D})_4$-brane, underlying a closed/open string duality, is argued to describe the Einstein vacuum in a low energy limit. We obtain topological de Sitter and Schwarzschild brane universe in six dimensions. The brane/anti-brane geometries are analyzed to explore some of their  characteristic and thermal behaviours in presence of the quantum effects. They reveal an underlying nine dimensional type 
IIA and IIB superstring theories on $S^1$.
\end{abstract}

\vfil\eject

\section{Introduction}
In the past years cosmic microwave background (CMB) has revealed the importance of a de Sitter vacuum in Einstein gravity. Astrophysical data from type Ia supernovae indicate an acceleration in cosmic expansion \cite{riess}. The cosmological limit from CMB leading to a small positive vacuum energy density motivates an intense research to explore de Sitter vacuum with renewed perspectives \cite{bousso-hawking}-\cite{paranjape}.

\sp
\noindent
Generically a de Sitter (dS) black hole is bounded by a cosmological horizon which makes it very different than an anti de Sitter (AdS) or an asymptotically flat black hole. The dS vacuum defines a maximally symmetric space. It has been conjectured that an asymptotic dS is bounded by a dS entropy. However an asymptotic dS does not possess a spatial infinity unlike to that in an asymptotic AdS. It poses a conceptual difficulty to realize the conjectured dS/CFT holographic duality \cite{hawking-maldacena-strominger, strominger-ds,klemm} between a quantum gravity on a $dS_{n+1}$ to a euclidean CFT on $S^n$.

\sp
\noindent
In particular an observer in a Schwarzschild de Sitter (SdS) black hole is bounded within a temporal phase. In addition a SdS universe is defined with a positive gravitational mass. A negative mass makes a SdS unphysical as it would describe a naked singularity. However a negative mass SdS may become sensible when formally identified with a topological de Sitter (TdS) black hole with a positive mass. The cosmological horizon in the SdS acts 
as an event horizon in the TdS which protects an observer in a space-like regime from hitting a curvature singularity. Very recently a negative mass SdS
black hole has been obtained in presence of a perfect fluid in Einstein gravity \cite{paranjape}. Importantly the energy-momentum tensor corresponding to a perfect fluid is argued to smear a point mass with a non-zero minimal length scale. As a result a curvature singularity at $r\rightarrow 0$ is not accessible for a negative mass SdS in Einstein vacuum. Thus a negative mass SdS gravity, coupled to fluid dynamics, becomes physical due to an intrinsic minimum length scale in the theory. A negative mass SdS in presence of a minimal length scale is characterized purely by a cosmological horizon. The disappearance of an event horizon in a negative mass SdS allows an observer to confine to a space-like regime. It provokes thought to imagine that a positive mass TdS across a cosmological horizon presumably describes a negative mass SdS in presence of a perfect fluid $T_{\mu\nu}$. 
Remarkably for a quantum primordial black hole, a (negative and positive) pair mass is identified with a pair of (electric and magnetic) non-linear charge square created in early universe at a Big Bang singularity.

\sp
\noindent
Interestingly the quantum effects as corrections to Einstein vacuum have been investigated in the folklore of theoretical high energy physics and cosmology. A microscopic black hole incorporates quantum effects at the beginning of the universe, where a curvature singularity is believed 
to exist classically. The quantum mechanical effect underlying an ultra-violet cut-off is believed to smear a point charge. Thus the microscopic black hole becomes free from a curvature singularity \cite{modesto} and may be viewed as a high energy trapped regime. A black hole in a quantum regime is believed to be governed in a superstring theory which perturbatively incorporates a quantum effect into its low energy effective string vacuum \cite{garfinkle-HS}. On the other hand, a quantum correction may as well be governed by a non-perturbative world underlying a strong-weak coupling duality in five dimensional superstring effective theories \cite{kar-maharana-panda}. A non-perturbative quantum effect, to a macroscopic black hole in a low energy string effective action, may explain the dark energy in our universe. The idea of a non-perturbative quantum correction to a perturbative (low energy) string vacuum is believed to provide a clue to a quintessence scalar presumably sourcing the dark energy. A 
quintessence may seen to describe a rapid growth in acceleration of our expanding universe underlying a hidden (or extra) fifth dimension.

\sp
\noindent
On the other hand Dirichlet (D) $p$-branes are believed to possess a potential tool to explore a new vacuum in a superstring theory. A $D_p$-brane is
charged under their Ramond-Ramond (RR) forms and they are non-perturbative objects propagating in a perturbative string theory \cite{polchinski}.
A space filling $D_9$-brane may seen to be described by the closed string modes. However the $D$-brane dynamics is primarily governed by an open string fluctuations at its boundary. Einstein gravity underlying a closed string is known to decouple from a $D$-brane world-volume at a critical value of a non-linear electric field. Thus a $D$-brane world volume is governed by a gauge theory with a flat metric. A global NS two form in a gauge invariant combination with an electromagnetic field has been shown to describe a nonlinear $U(1)$ gauge symmetry  \cite{seiberg-witten}. The significance of non-linear, electric and magnetic, charges have been explored to describe various near horizon deformations on a $D$-brane \cite{gibbons}-\cite{grezia-esposito-miele}.

\sp
\noindent
Furthermore a $4D$ de Sitter vacuum has been realized in a low energy string theory \cite{kklt}. It was remarkably constructed by lifting an AdS minimum by an addition of anti $D_3$-brane in a type IIB superstring theory. Subsequently dS space has been worked out in a superstring theory by adding fluxes of form fields within the $D_7$-branes \cite{burgess-kallosh-quevedo}. These constructions allow to view a dS via a spontaneously broken supersymmetric vacuum in $4D$. Very recently the dS vacua in type IIB superstring theory has been argued with a careful control of higher order curvature corrections \cite{keshav-tatar}.

\sp
\noindent
In the context we recall a pair production mechanism, underlying a particle and an anti-particle, at a black hole horizon has been established as a powerful tool to address some of the quantum gravity effects in the folklore of theoretical physics \cite{hawking,hawking-gibbons,mottola}. The field theoretic idea was generically applied in presence of a non-trivial background metric leading to a black hole. In particular, a photon in a $U(1)$ gauge theory is known to produce an electron-positron pair at a black hole event horizon which in turn is believed to explain the Hawking radiation phenomenon. Interestingly a pair of $(D{\bar D})$-brane creation mechanism at a cosmological horizon has been explored to address some aspects of inflation in string cosmology \cite{burgess-majumdar,majumdar-davis}. 

\sp
\noindent
In the recent past, an effective five dimensional torsion curvature underlying vacuum created pair of $(D{\bar D})_3$-brane has been constructed in a second order formalism \cite{spsk-JHEP}-\cite{kpss}. In particular, a two form quanta in a $U(1)$ gauge theory on a $D_4$-brane was argued to produce a pair of $(D{\bar D})_3$-brane at a cosmological horizon which turns out to describe a Big Bang singularity. An underlying electric non-linear $U(1)$ charge in the effective curvature formalism incorporates an extended particle/anti-particle pair creation in the quantum theory. The assertion for a pair of (non-fundamental) strings may also be confirmed with a local two form on a $D_p$-brane. The stringy nature further reassures an underlying quantum gravity phenomenon in Yang-Mills theory \cite{thooft,kar-maharana,hanada}. Interestingly a string/five-brane duality in the underlying ten dimensions may be exploited to obtain a $D_5$-brane universe in the formalism. 

\sp
\noindent
We may recall that a BPS $D_5$-brane is purely governed by a two form world-volume CFT in presence of a background open string metric (see fig-1). The vacuum created pair of $(D{\bar D})_4$-brane at a cosmological horizon has been argued at the expense of a local two form on a BPS $D_5$-brane (see fig-2). A pair of BPS brane and an anti BPS brane breaks the supersymmetry and describes a non-BPS configuration. The RR charge of a vacuum created $D_4$-brane is nullified by the opposite RR charge of ${\bar D}_4$-brane within a pair. Nevertheless they possess their origin in a higher dimensional BPS brane which carries a RR charge of a five brane or its dual a $D$-string. The vacuum created brane/anti-brane geometries break the supersymmetry in presence of an extra sixth dimension transverse to the brane/anti-brane world-volumes. In other words an extra transverse dimension to 
the vacuum created pair ($D_4$-brane or an anti ${\bar D}_4$-brane world-volume) may seen to incorporate the low energy closed string modes into a BPS brane in the formalism. Thus a vacuum created $D_4$-brane geometry may be approximated by a $5D$ Einstein vacuum in presence of a typical $D_4$-brane. 

\sp
\noindent
Generically a higher form is Poincare dual to a one form on an appropriate $D_p$-brane for $p>3$. Thus a higher form theory in presence of a background dS metric may describe a higher dimensional vacuum pairs such as a membrane/anti-membrane, a three brane/anti-brane and other higher dimensional brane pairs. It may be recalled that a space filling brane ($D_9$-brane) in type IIB superstring theory is indeed described by the closed string modes there. The $U(1)$ gauge invariance, underlying a two form on a $D_4$-brane, has been exploited and was shown to incorporate a metric fluctuation in an effective curvature formalism \cite{spsk-JHEP,spsk-NPB-P}. In addition a vacuum created $(D{\bar D})_3$ was shown to be described with an extra (transverse) dimension within a pair and is viewed as a space filling $D_4$-brane \cite{spsk-PRD}. Interestingly the vacuum geometries on a pair of $(D{\bar D})_3$-brane has been identified with a low energy (perturbative) string vacuum in presence of a non-perturbative quantum correction sourced by a lower dimensional $D_3$-brane or an anti $D_3$-brane \cite{spsk-NPB1,spsk-NPB2}. 

\sp
\noindent
At this point it is worth mentioning that a background NS two form is known to introduce a (string theoretic) torsion in the low energy string effective action \cite{candelas-HS,freed}. However the effective curvature formalism deals with 
a world-volume torsion on a $D_4$-brane which a priori has no relevance to a string or NS-torsion. We may recall that a world-volume torsion in an effective curvature formalism may be identified with a two form or its Poincare dual (non-linear one form) gauge field on a $D_4$-brane. Intuitively an induced NS-torsion may be realized via a pull-back in a world-volume of an anti $D_4$-brane which serves as a background to a $D_4$-brane in the formalism \cite{pssk1}-\cite{kpss}. In fact our analysis, for a $D_5$-brane in the paper, re-assures an underlying vacuum pair of $(D{\bar D})_4$-brane.

\sp
\noindent
In the paper we revisit an effective torsion curvature formalism underlying a two form world-volume dynamics in a $D_4$-brane \cite{spsk-JHEP,spsk-NPB-P}. We  work out an effective curvature in six dimensions sourced by a two form $U(1)$ gauge theory on a $D_5$-brane. Six dimensional curvature in the formalism becomes special due to the underlying CFT on a $D_5$-brane. In principle a higher dimensional construction would involve a higher form on a $D_p$-brane which in turn is believed to enhance our knowledge on the conjectured M-theory in eleven dimensions. For instance the closed string modes in a space filling $D_9$-brane in type IIB superstring theory would be described by a vacuum created pair of $(D{\bar D})_8$-brane in the formalism. An appropriate higher form dynamics on a space filling brane in presence of a background anti $D_9$-brane may provoke thought to look for a space filling $9$-brane in $M$-theory \cite{pssk-jaat}. However we restrict the formalism to six dimensions in the paper which may 
be viewed via a non-coincident pair of $(D{\bar D})_5$-brane with a global NS two form on an anti brane and a local two form on a brane. In the case a global NS two form would lead to open string metric on an anti $D_5$-brane which in turn describes a background metric in the $D_5$-brane world-volume theory. Thus our starting point may be enumerated by a two form $U(1)$ gauge theory in presence of a background (open string) metric on a $D_5$-brane. We compute the energy-momentum tensor in the six dimensional CFT to show the existence of a nontrivial metric in the effective torsion curvature underlying a vacuum created pair of $(D{\bar D})_4$-brane. A two form quanta in a CFT is argued to produce a pair of brane and anti-brane at a cosmological horizon. 

\sp
\noindent
Interestingly the emergent gravity scenario is shown to describe a low energy (perturbative) string vacuum in $6D$ with a (non-perturbative) quantum correction by a lower ($p<5$) dimensional $D_p$ brane or an anti brane in the formalism. A closed string 
exchange between a pair of $(D{\bar D})_4$-brane, underlying a closed/open string duality, may be helpful to realize the Einstein vacuum in a low energy limit. We obtain topological de Sitter and Schwarzschild brane universes in six dimensions. The brane/anti-brane universes are analyzed for their characteristic properties. They are qualitatively argued to be cosmologically created as a negative and positive mass pairs from a vacuum across a horizon. Their thermal behaviours are analyzed in presence of the quantum effects. It is argued that a quantum geometries undergo intermittent geometric transitions underlying various lower dimensional pairs of brane/anti-brane. They lead to a thermal equilibrium by losing energy via Hawking radiations and describe a stable brane universe at low energy. Their potentials reveal an underlying nine dimensional type IIA and IIB superstring theories on $S^1$.

\sp
\noindent
We plan the paper as follows. We begin with a moderate introduction in section 1. A geometric torsion curvature formalism in six dimensions is briefly discussed in section 2 to obtain some of the de Sitter causal patches on a vacuum created pair of $(D{\bar D})_4$-brane. The quantum corrections, underlying various lower dimensional $D$-branes, to the low energy (perturbative) string vacuum in six dimensions are worked out to describe a plausible non-perturbative quantum gravity in the formalism. We obtain TdS and SdS branes in $6D$,  analyze their characteristics and explore some of their thermal aspects leading to stable brane universe at equilibrium in section 3. We conclude the paper with some remarks in section 4. 

\section{Torsion geometries in ${\mathbf{6D}}$}
\subsection{Higher form gauge theory on a ${\mathbf{D_p}}$-brane}
A $D_p$-brane carries a Ramond-Ramond (RR) charge and is established as a non-perturbative dynamical object in a ten dimensional type IIA or IIB superstring theories \cite{polchinski}. In particular a $D_5$-brane is governed by a supersymmetric gauge theory on its six dimensional world-volume in a ten dimensional type IIB superstring theory. However we restrict to the bosonic sector and begin with the $U(1)$ gauge dynamics, in presence of a constant background metric $g_{\mu\nu}$, on a $D_5$-brane. A linear one form dynamics is given by
\be
S_{\rm A}= -{1\over{4C_1^2}}\int d^6x\ {\sqrt{-g}}\ F^2\ ,\label{gauge-1}
\ee 
where $C_1^2=(4\pi^2g_s)\alpha'$ denotes the gauge coupling. A non-linear $U(1)$ gauge symmetry is known to be preserved in a one form theory in presence of a constant NS two form on a $D$-brane. An electric non-linear charge is known to govern an effective gravity, underlying an open string metric $G_{\mu\nu}^{(NS)}$, on a $D$-brane \cite{seiberg-witten}. The six dimensional non-linear gauge dynamics may be approximated by a Dirac-Born-Infeld (DBI) action which describes a non-linear one form ${\cal A}_{\mu}$. It is given by
\be
S_{\cal A}= -T_{D}\int d^6x\ {\sqrt{-\left ( g + B + {\bar F}\right )}} \ .\label{gauge-2}
\ee 
On the other hand the Poincare duality ensures that the world-volume gauge dynamics on a $D_p$-brane for $p>3$ may appropriately be described by a $m$-form for $m>1$. For instance the $U(1)$ gauge theory on a $D_5$-brane may also be re-expressed in terms of a three form $B_3$. In principle a higher ($m$) form gauge theory on a higher dimensional $D_p$-brane may be explored to address a non-perturbative quantum gravity formulation leading to an eleven dimensional M-theory. Generically a higher form gauge theory underlying a background de Sitter vacuum on a higher dimensional $D_p$-brane may describe a number of vacuum pairs such as: a membrane/anti-membrane, a three or higher dimensional brane/anti-brane universes.

\sp
\noindent 
In the context two of the authors (RK and SK) in a collaboration have revisited an effective curvature formalism underlying a $U(1)$ gauge theory on a $D_4$-brane in the recent past \cite{psskk}. A five dimensional geometric torsion dynamics was exploited to describe the creation of a pair of 
$(D{\bar D})_3$-brane universe. The quitessence axion dynamics on an anti $D_3$-brane was argued to describe the space-time curvature on a $D_3$-brane universe. A quintessence scalar is known to incorporate a variable vacuum energy density and is believed to describe the conjectured dark energy in our universe. In fact the effective torsion curvature in five dimensions described by a vacuum created pair of $(D{\bar D})_3$-brane at a cosmological horizon was developed by one of the authors (SK) in a collaboration \cite{spsk-JHEP,spsk-PRD,spsk-NPB-P,spsk-NPB1}. The second order formalism was used to address AdS brane, de Sitter brane, Kerr brane and Kerr-Newman brane, Reissner-Nordstrom brane and Schwarzschild brane universes and their tunnelings \cite{spsk-NPB2,pssk-jaat,pssk1,pssk2,kpss}. 

\sp
\noindent
For simplicity we consider a two form $U(1)$ gauge dynamics on a $D_5$-brane in presence of a background open string metric. It  was shown that a global NS two form underlying the open string metric can source a pure de Sitter vacuum. A vacuum pair of $(D{\bar D})_4$-brane has been argued to be created, at the cosmological horizon in a background de Sitter, by a two form quanta in a non-linear $U(1)$ gauge theory on a $D_5$-brane. The pair production mechanism is inspired by the novel idea of Hawking radiation at the event horizon of a semi-classical black hole by a photon in a $U(1)$ gauge theory \cite{hawking}.

\sp
\noindent
A vacuum created pair $(D{\bar D})_4$ is described with an extra sixth transverse dimension between a brane and an anti-brane and may be viewed as a space filling $D_5$-brane. It may be recalled that a space filling brane ($D_9$-brane) in type IIB superstring theory is indeed described by the closed string modes there. Moreover the $U(1)$ gauge invariance of the two form on a $D_4$-brane has been exploited to enforce the metric fluctuations in the five dimensional effective curvature formalism \cite{spsk-JHEP}. It was shown that the brane universe may be viewed through a low energy perturbative string vacuum in presence of a quantum correction underlying a (non-perturbative) $D$-brane or an anti $D$-brane. 

\sp
\noindent
Now we consider a two form $U(1)$ gauge theory on a $D_5$-brane world-volume in presence of a background (open string) metric $G_{\mu\nu}^{(NS)}$. 
See Figure 1 for a schematic set-up in a gauge theory. The action may be given by
\be
S_{\rm B}=- {1\over{12C_2^2}}\int d^6x\ {\sqrt{-G^{(NS)}}}\;\ H_{\mu\nu\lambda}H^{\mu\nu\lambda}\ ,\label{gauge-3}
\ee
where $C_2^2=(16\pi^4g_s){\alpha'}^2$ denotes a gauge coupling. The local degrees in two form, on a $D_5$-brane, shall be exploited to construct an effective space-time curvature scalar ${\cal K}$ in a second order formalism \cite{spsk-JHEP}. Importantly a two form gauge theory in $6D$ retains  conformal invariance at the classical level.
\subsection{Two form ansatz}
A two form ansatz in a gauge theory on a $D_4$-brane has been shown to describe a five dimensional geometric torsion in a second order formalism \cite{spsk-JHEP}. In the case a two form on a $D_5$-brane may seen to describe a six dimensional effective curvature ${\cal K}$. Thus a geometric torsion ${\cal H}_3$ in the formalism may alternately be viewed via a vacuum created pair of $(D{\bar D})_4$-brane. The five dimensional brane world-volumes, within a vacuum created pair, are separated by an extra (sixth) transverse dimension. The scenario may be described by an irreducible scalar curvature ${\cal K}$, underlying the $U(1)$ gauge theories on a $D_5$-brane, on $S^1$. 
\begin{figure}
\centering
\mbox{\includegraphics[width=.5\linewidth,height=0.27\textheight]{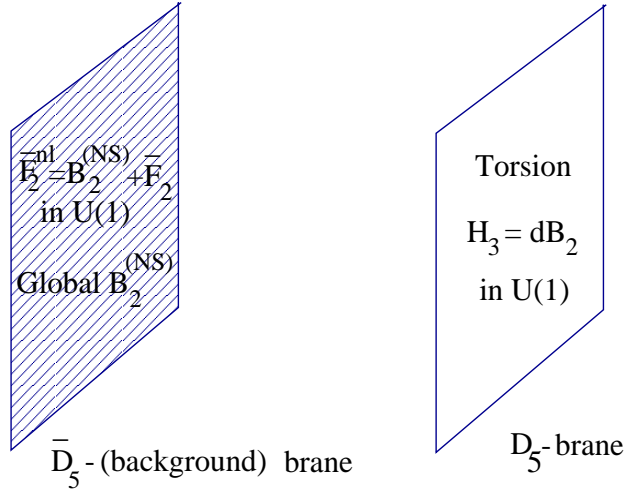}}
\caption{\it {Schematic set up describing a two form $U(1)$ gauge theory on a $D_5$-brane in presence of a background anti $D_5$-brane.}}
\end{figure}

\sp
\noindent
Generically a vacuum created $D_p$-brane in the formalism is linked to an anti $D_p$-brane through an extra transverse dimension. In particular we consider a global NS two form $B_2^{(NS)}$ on an anti $D_5$-brane and a dynamical $B_2$ on a $D_5$-brane underlying a global scenario. A global NS two form is known to describe an effective open string metric on an anti $D_5$-brane which turns out to be a background metric on a $D_5$-brane in the set up. The gauge field ansatz may be given by
\bea
&&B_{\psi t}^{(NS)}\ =\ B_{\psi r}^{(NS)}={{b}\over{(2\pi\alpha')^{1/2}}}\ ,\nonumber\\
{\rm and}\quad &&B_{\psi\phi}\ =\ {{P^3}\over{(2\pi\alpha')^{3/2}}}\ (\sin^2\psi\ \cos\theta)\ ,\label{D5-twoform-1}
\eea
where $(b,P)>0$ are constants. They shall be identified with the conserved quantities defined in an asymptotic regime underlying an effective brane geometry. The global NS two form, in a gauge invariant combination with an electromagnetic field, is known to govern the DBI dynamics on an anti $D_5$-brane. The open string effective metric is described by an electric non-linear charge $b$ which in turn describes an effective de Sitter causal patch. The other parameter $P$ incorporates a dynamical two form into a $D_5$-brane in presence of an effective de Sitter background metric. A nontrivial component of the field strength is worked out to yield:
\be 
H_{\psi\theta\phi} =\frac{P^3}{(2\pi\alpha')^2}\,({\sin}^2\psi \sin\theta)\ .\label{D5-twoform-2}
\ee
An electric-magnetic self-duality in a six dimensional world-volume may be explored to interpret $P$ as a magnetic as well as an electric non-linear $U(1)$ charge. The parameter $P$ incorporates a gauge theoretic torsion and satisfies the two form field equation:
\bea
&&{\nabla}_{\lambda}H^{\lambda\mu\nu}\ =\ 0\nonumber\\
{\rm or}\qquad&&\partial_{\lambda}H^{\lambda\mu\nu}\ +\ {1\over2}\Big ( g^{\alpha\beta}\partial_{\lambda}\ g_{\alpha\beta}\Big )
H^{\lambda\mu\nu}\ =\ 0\ , \label{D5-twoform-21}
\eea
where $g_{\mu\nu}$ is a flat metric defined with a spherical symmetry. It is given by
\be 
ds^2=- dt^2 + dr^2 + r^2 d\beta^2+r^2_{\beta} d\psi^2+r^2_{\beta}\sin^2\psi
d\theta^2+r^2_{\beta}\sin^2\psi\sin^2\theta d\phi^2\label{D5-twoform-2}
\ee
$$\rm{where}\quad r_\beta\ =\ r\sin\beta\ .\qquad\qquad\qquad\qquad\quad\;\ \qquad\qquad\qquad\qquad\qquad\qquad {}$$
\noindent
The angular coordinates are defined with $(0\leq\psi\leq\pi)$, $(0\le\beta\le\pi)$, $(0\le\theta\le\pi )$ and $(0\le\phi< 2\pi )$.
They describe $S^4$ symmetric vacuum configuration and the line-element is given by
\bea
d\Omega_4^2&=&d\beta^2\ +\ \sin^2\beta\ \left ( d\psi^2\ +\ \sin^2\psi d\theta^2\ +\ \sin^2\psi \sin^2\theta\ d\phi^2\right )\ ,\nonumber\\ 
&=&d\beta^2\ +\ \sin^2\beta\ d\Omega_3^2\ .\label{D5-twoform-3}
\eea
For simplicity we identify some of the $S^2$ symmetric line-elements within $S^4$ and they are given by
\bea
&&d\Omega_\psi^2\ =\ d\psi^2\ +\ \sin^2\psi\ d\theta^2\ , \nonumber\\
&&d\Omega_\theta^2\ =\ d\theta^2\ +\ \sin^2\theta\ d\phi^2\  \nonumber\\
{\rm and}\quad&&d\Omega_\beta^2\ =\ d\beta^2\ +\ \sin^2\beta\ d\phi^2 \ .\label{D5-twoform-4}
\eea
\subsection{Geometric torsion}
A geometric torsion sourced by (global NS and local) two forms in an effective curvature formalism may be constructed from the underlying perturbative $U(1)$ gauge theory on a $D_5$-brane. A global NS two form on an anti-brane is known to couple perturbatively to a gauge theoretic torsion $H_3$ on a brane within a pair \cite{spsk-JHEP}. It may be given by
\bea 
{\cal H}_{\mu\nu\lambda}&=& {\cal D}_{\mu}B_{\nu\lambda}\ +\ {\cal D}_{\nu}B_{\lambda\mu} \ +\ {\cal D}_{\lambda}B_{\mu\nu} \nonumber\\
&=&3\nabla_{[\mu}B_{\nu\lambda ]}\ +\ 3{{\cal H}_{[\mu\nu}}^{\alpha}B^{\beta}_{(NS)\ \lambda ]}\ g_{\alpha\beta}\nonumber\\ 
&=&H_{\mu\nu\lambda}\ +\ \left ( H_{\mu\nu\alpha}B^{\alpha}_{(NS)\ \lambda} + {\rm cyclic\ in}\ \mu,\nu,\lambda\right )\ +\ 
{\cal{O}}\left (B_{(NS)}^2\right )\ .\label{D5-twoform-5}
\eea
It may be noted that a geometric torsion ${\cal H}_3$ is defined with a modified covariant derivative ${\cal D}_{\mu}$. Interestingly ${\cal H}_3$
incorporates a constant NS two form as a perturbative coupling to a dynamical two form in the world-volume gauge theory. 

\sp
\noindent
At this point we digress and recall that a $B^{(NS)}_2$ couples to an electromagnetic field $F_2$ and forms a $U(1)$ gauge invariant combination on a $D_p$-brane. A global mode can not be gauged away from the theory and hence the $U(1)$ gauge invariant field strength becomes nonlinear on a $D_p$-brane. Thus a $B^{(NS)}_2$ coupling to a gauge field strength $H_3$ is believed to nurture the non-linearity in a $U(1)$ gauge theory. Under a two form $U(1)$ gauge transformation a Lorentz scalar ${\cal H}^2$ has been shown to be gauge invariant in presence of an emerging notion of a metric fluctuation \cite{spsk-JHEP,psskk}. Most importantly a metric dynamics is elegantly governed by the Einstein-Hilbert action which is based on a second order formulation. The Einstein metric under a Weyl scaling may be identified with a string metric whose dynamics is ensured by the vanishing of 
$\beta$-function equations. Generically a string metric along with a NS two form and a dilaton is known to be described in a (low energy) superstring effective action. In other words the coupling of a constant NS two form enforces a string charge perturbatively into a two form gauge theory on a $D_p$-brane world-volume which may hint at a second order formalism underlying an effective curvature \cite{spsk-JHEP,spsk-PRD,spsk-NPB-P,spsk-NPB1,spsk-NPB2}.
\begin{figure}
\centering
\mbox{\includegraphics[width=.5\linewidth,height=0.27\textheight]{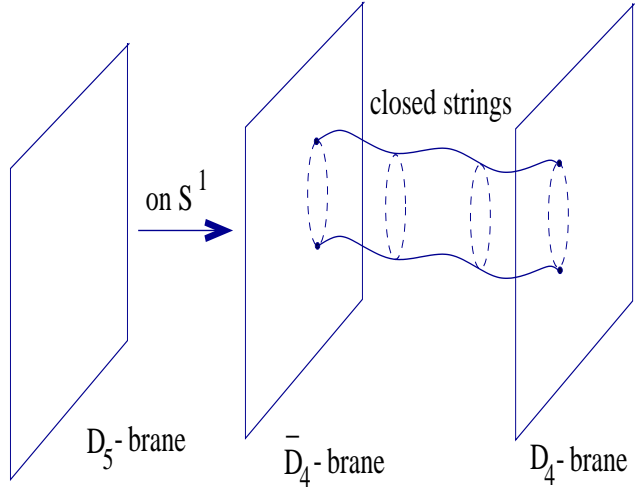}}
\caption{\it A vacuum created pair of $(D{\bar D})_4$-brane may seen to exchange closed strings underlying the open/closed string duality. The presence of an extra dimension may allow one to view the six dimensional brane universe on $S^1$.}
\end{figure}

\sp
\noindent
A commutator, of covariant derivatives, was worked out to explore the possibility of an effective curvature in a second order formalism.
For a Minkowski metric $g_{\mu\nu}$, the commutator simplifies to describe an effective curvature tensor of order four. It is given by
\be
\Big [ {\cal D}_{\mu}\ ,\ {\cal D}_{\nu} \Big ]A_{\lambda}=\ {{\cal K}_{\mu\nu\lambda}{}}^{\rho}A_{\rho}\ ,\label{geometricT-3}
\ee 
where
\be
{{\cal K}_{\mu\nu\lambda}{}}^{\rho}\ \equiv\ \partial_{\mu}{{\cal H}_{\nu\lambda}}^{\rho}\ -\ \partial_{\nu} 
{{\cal H}_{\mu\lambda}}^{\rho}\ +\ {{\cal H}_{\mu\lambda}}^{\sigma}{{\cal H}_{\nu\sigma}}^{\rho}\ -\ {{\cal H}_{\nu\lambda}}^{\sigma}{{\cal H}_{\mu\sigma}}^{\rho}\ .\label{gauge-12}
\ee
The space-time curvature tensor is anti-symmetric under an interchange of a pair of indices, $i.e.\ {\cal K}_{\mu\nu\lambda\rho}
=-{\cal K}_{\nu\mu\lambda\rho}=-{\cal K}_{\mu\nu\rho\lambda}$. The effective curvature tensor incorporates the dynamics of a geometric torsion in six 
dimensions. For a non propagating torsion, in a gauge choice, the effective curvature tensor reduces to the Riemannian tensor: 
${\cal K}_{\mu\nu\lambda\rho}\rightarrow R_{\mu\nu\lambda\rho}$. The gauge choice ensures a decoupling of a dynamical two form in a perturbative gauge theory which in turn dissociates a non-perturbative ($D$-brane) correction from the low energy string vacuum. Other relevant curvature tensors are worked out to yield:
\bea
&&{\cal K}_{\mu\nu}\ \equiv\ -\ \left (\partial_{\lambda}{{\cal H}^{\lambda}}_{\mu\nu} +
{{\cal H}_{\mu\rho}}^{\lambda}{{\cal H}_{\lambda\nu}}^{\rho}\right )
\nonumber\\
{\rm and}\quad&&{\cal K}\ \equiv\ -\ {\cal H}_{\mu\nu\lambda}{\cal H}^{\mu\nu\lambda}\ .\label{gauge-17}
\eea
The effective curvature scalar ${\cal K}$ in six dimensions underlies a two form $U(1)$ gauge theory on a $D_5$-brane. Thus the world-volume dynamics on a generic $D_p$-brane for $p>3$ may be described by a geometric torsion dynamics in a second order effective curvature formalism. It is worth mentioning that an effective torsion curvature takes into account a string charge, sourced by a background $B^{(NS)}_2$, in addition to a non-linear $U(1)$ gauge theory on a $D_5$-brane. Thus the effective torsion curvature formalism on a $D_p$-brane may be viewed via a vacuum created pair of $(D{\bar D})_{p-1}$ brane under the exchange of closed strings in a dual channel. See Figure 2 for a schematic scenario. An extra transverse dimension between a $D_{p-1}$ brane and an anti $D_{p-1}$-brane signals the presence of Einstein gravity in the formalism. In other words a $D_5$-brane vacuum in presence of a geometric torsion may equivalently be described by a low energy $6D$ string vacuum in presence of a lower ($p<5$) 
dimensional $D_p$-brane or an anti $D_p$-brane correction \cite{spsk-PRD,spsk-NPB2,psskk,pssk1,pssk2}. It was argued that a lower dimensional pair of $(D{\bar D})_{p-1}$ brane universe is created at a cosmological horizon by a two form quanta in a $U(1)$ gauge theory on a $D_p$-brane with a de Sitter background. A geometric torsion incorporates a spin angular momentum into the vacuum created pair. A spinning motion on a vacuum created $D_p$-brane is in opposite direction to that on anti $D_p$-brane. It is remarkable to note that the effective curvature scalar ${\cal K}$ takes into account a non-perturbative correction into a perturbative (low energy) string vacuum. The effective world-volume dynamics in the case may be described by 
\be
S= {1\over{3C_5^2}}\int d^6x {\sqrt{-G^{(NS)}}}\;\;\Big ( {\cal K}\ -\ \Lambda\Big )\ ,\label{gauge-11}
\ee
where $C_5$ signifies an appropriate coupling constant underlying a $D_5$-brane tension and $\Lambda$ denotes a cosmological constant. The invariant volume signifies the presence of an open string effective metric determinant $G^{(NS)}$ in the formalism. Furthermore a geometric torsion in the effective action may seen to be sourced by an energy-momentum tensor $T_{\mu\nu}$. The trace of $T_{\mu\nu}$ is computed to yield $T_{\mu}^{\mu}=\Lambda$. It shows that the conformal invariance at a classical level may seen to be broken by the presence of a cosmological constant. With a gauge choice, ${\Lambda}=\Big (3(\pi\alpha')^{-1}+{\cal K}\Big )$, the $T_{\mu\nu}$ may seen to source an emergent metric. Explicitly it is given by
\be
T_{\mu\nu}\ =\ {1\over{2\pi\alpha'}}\left ( G^{(NS)}_{\mu\nu}\ +\ \left ( C-{1\over4}\right )\ {\bar{\cal H}}_{\mu\lambda\rho} {{\cal H}^{\lambda\rho}}_{\nu}\right )\ .\label{gauge-31}
\ee
The emergent metric underlying a geometric torsion dynamics (\ref{gauge-11}) is worked to yield:
\bea
{\hat G}_{\mu\nu}&=&\left ( g_{\mu\nu}\ -\ B^{(NS)}_{\mu\lambda}g^{\lambda\rho}B^{(NS)}_{\rho\nu}\ +\ C\ {\bar{\cal H}}_{\mu\lambda\alpha}g^{\lambda\rho}
g^{\alpha\beta}{\cal H}_{\rho\beta\nu}\right )\nonumber\\
&=&\left ( G^{(NS)}_{\mu\nu}\ +\ C\ {\bar{\cal H}}_{\mu\lambda\alpha}g^{\lambda\rho}g^{\alpha\beta}{\cal H}_{\rho\beta\nu}\right )\ .\label{emetric}
\eea
A geometric torsion correction to an open string effective metric ensures an extra transverse dimension between a vacuum created pair of $(D{\bar D})_4$-brane in the formalism. The hidden dimension to a brane universe may provide a clue to describe a quintessence axion on a lower dimensional 
anti brane universe \cite{psskk,pssk1,pssk2}. 

\sp
\noindent
Now we work out the geometric torsion from the ansatz (\ref{D5-twoform-1}) on a $D_5$-brane. Its non-trivial components are given by
\bea
&&{{\cal H}_{\theta\phi}}^{\psi}\ =\ (2\pi\alpha')^{-1}\ {{P^3}\over{r^2_{\beta}}}\ (\sin^2\psi \sin \theta)\nonumber\\
{\rm and}\quad&&{{\cal H}_{\theta\phi}}^t\ =\ -{{\cal H}_{\theta\phi}}^r\ =\ -(2\pi\alpha')^{-3/2}\ {{bP^3}\over{r^2_\beta}} (\sin^2\psi \sin \theta)\ .\label{D5-twoform-6}
\eea
It shows that a global NS two form can generate an electric field from a magnetic field and vice-versa. 
As a result a magnetic non-linear charge has been shown to be generated from an electric point charge without a magnetic monopole \cite{pssk2}.
We set $C=\pm(1/2)$ in the emergent metric expression (\ref{emetric}) on a $D_5$-brane. It may also be argued to be sourced by ${\hat T}_{\mu\nu}$ in (\ref{gauge-31}) for $C=\mp(1/4)$). It is given by
\be 
{\hat G}_{\mu\nu} = \left (g_{\mu\nu}\ -\ B_{\mu\lambda}^{(NS)}g^{\lambda\rho}B_{\rho\nu}^{(NS)}\right )_{\rm BG}\ \mp\ \frac{1}{2}{\bar{\cal H}}_{\mu\lambda\alpha}\ g^{\rho\alpha} g^{\lambda\sigma}\ {\cal H}_{\rho\nu\sigma}\ ,\label{D5-twoform-7}
\ee
where the subscript ${\rm BG}$ denotes a background metric on a $D_5$-brane. It is the open string metric sourced by a global NS two form on an anti $D_5$-brane which serves as a background to a $D_5$-brane in the formalism. The emergent metric components are worked out with $(2\pi\alpha')=1$ to yield:
\bea 
&&{\hat G}_{tt}\ =\ -\ \left (1\ -\ {{b^2}\over{r^2_\beta}}\ \pm\ {{b^2P^6}\over{r^8_\beta}} \right )\ ,\quad {\hat G}_{rr}= \left (1\ +\ {{b^2}\over{r^2_\beta}}\ \mp\ {{b^2P^6}\over{r^8_\beta}}\right )\ ,\nonumber\\
&&{\hat G}_{\psi\psi}=\left ( 1\ \mp\frac{P^6}{r^6_\beta}\right )r_{\beta}^2\ ,\;\;\; {\hat G}_{\theta\theta}={\hat G}_{\psi\psi} \sin^2\psi\ ,
\;\; {\hat G}_{\phi\phi}= {\hat G}_{\psi\psi}\sin^2\psi\sin^2\theta\ ,\nonumber\\
&&{\hat G}_{\beta\beta}\ =\ r^2\ , \quad {\hat G}_{tr}\ =\ 
{{b^2}\over{r^4_{\beta}}}\ {\hat G}_{\psi\psi}\quad {\rm and}\quad {\hat G}_{t\psi}\ =\ {\hat G}_{r\psi}=\mp\ \frac{bP^6}{r^6_\beta}\ .\label{D5-twoform-8}
\eea
The emergent patches on a $D_5$-brane are indeed sourced by the energy momentum tensor (\ref{gauge-31}). Thus they are sourced by a two form $U(1)$ gauge theory in a six dimensional world-volume. Alternately they reassure the propagation of geometric torsion in a six dimensional effective space-time curvature constructed in a second order formalism. In a certain brane window, the geometric patches may formally be identified with a black hole geometry underlying a brane universe.

\sp
\noindent
Interestingly the constructed geometric patches in a low energy limit shall be seen to describe an established Einstein vacuum in section 2.5. Generically the limit may be described by a gauge choice which in turn corresponds to a non-propagating torsion in the effective curvature formalism \cite{spsk-NPB1,spsk-NPB2}. In fact a Riemannian curvature is reassured by a non-propagating torsion \cite{spsk-PRD}. The local degrees of a two form freeze in the gauge theory in a low energy limit. Then the reduced brane geometry is purely governed by the background (constant) NS two form. This in turn identifies the reduced vacuum with that in a low energy string effective action. Neverthless the full geometries (\ref{D5-twoform-8}) through a certain brane window may seen to describe a typical $6D$ gravitational black hole in a low energy string theory in presence of a lower ($p<5$) dimensional $D_p$-brane or ${\bar D}_p$-brane. A BPS brane is shown to incorporate a gauge theoretic (quantum) correction into the stringy vacuum. It is argued that an effective torsion curvature in a six dimensional world-volume underlies a vacuum created pair of $(D{\bar D})_4$-brane in presence of an extra transverse dimension between them. Thus a geometric correction by a BPS $D_4$-brane or an anti BPS ${\bar D}_4$-brane, to the low energy string (Einstein) vacuum, is essentially governed by a local two form leading to a torsion $H_3$ in a $U(1)$ gauge theory. An analogous notion in five dimensions, leading to some of the nontrivial vacua, has been discussed in ref.\cite{pssk2}. Thus the emergent metric patches in the paper correspond to some of the (nonperturbative quantum) vacua in type IIA or IIB superstring theory on $S^1$.

\sp
\noindent
In a special angular slice, $i.e.$ for $\beta=\pi/2$, the metric components reduce to that obtained by one of the authors (SK) in a collaboration \cite{spsk-JHEP}. Nevertheless an arbitrary $\beta$-coordinate generalizes the vacuum to a higher dimension which possess an underlying CFT on a $D_5$-brane. An qualitative analysis further assures that a two form vacuum creats a pair of $(D{\bar D})$-string at the horizon which
may formally be identified with a pair of $(D{\bar D})_4$-brane under a duality in a type IIA or IIB superstring theory on $S^1$. This in turn justifies the significant role of a $D_5$-brane in the formalism. The torsion curvature formalism in the paper is a nontrivial generalization of that 
obtained on a $D_4$-brane. In addition the results in the paper explore certain new aspects such as the vacuum created pair of (positive and negative) mass in the near horizon geometry.

\sp
\noindent
On the other hand a three form potential (Poincare dual to a one form gauge field) on a $D_5$-brane may play a significant role to incorporate some of the quantum non-perturbative effects into the Einstein gravity. The role of a higher form in an appropriate world-volume theory may provide a clue to the conjectured M-theory in eleven dimensions. However for simplicity, we restrict to a two form gauge dynamics on a $D_5$-brane in presence of a stringy black hole background in the paper. Interestingly an effective torsion curvature in five dimensions \cite{spsk-JHEP,spsk-NPB-P} has been explored to address the cosmological origin of a brane or anti-brane universe at a Big Bang singularity. Subsequently the formalism is used to explain a pair production of a four dimensional (dS and AdS) brane and an anti brane universe underlying Einstein gravity in ref.\cite{spsk-PRD}. The torsion curvature in five dimensions were analyzed to obtain Kerr family of solutions underlying some of the string vacua in ref.\cite{spsk-NPB1,spsk-NPB2}. Importantly the torsion formalism was explored to address some aspects of quintessence axionic scalar dynamics \cite{psskk,pssk1}. The notion of fifth essence is indeed supported by the existance of a fifth transverse dimension between a vacuum created pair of $(D{\bar D})_3$-brane. In fact, quintessence is known to be a potential candidate to explain the observed acceleration and expansion of our universe which in turn is believed to source the conjectured dark energy.

\subsection{Weyl scaling: de Sitter causal patches}
In this section we perform an appropriate Weyl (conformal) transformation of the emergent metric (\ref{D5-twoform-8}) to access the de Sitter vacua on a vacuum created $D_5$-brane. The Weyl scaling may be given by
\be
{\hat G}_{\mu\nu}= {{b^2}\over{r^2_{\beta}}} G_{\mu\nu}\ .\qquad\qquad {}\label{dS-1}
\ee
In a brane window, $i.e.\ b>r_{\beta}>P$ with ($b^4>>r_{\beta}^4$ and $r_{\beta}^{12}>>P^{12}$), the conformal factor ensures an ultra high energy regime for the transformed vacua. A priori the emergent quantum geometries underlying a $D_5$-brane may be given by
\bea
ds^2&=&\left (1-{{r^2_{\beta}}\over{b^2}}\ \mp\ {{P^6}\over{r^6_\beta}}\right ) dt^2\ +\  
\left (1 -{{r^2_{\beta}}\over{b^2}}\ \pm\ {{P^6}\over{r^6_\beta}}\right )^{-1} dr^2 \nonumber\\
&&-\ {{2r_{\beta}}^2\over{b}}\left ( dt + dr \right )d\psi\ +\ {{r^2r^2_\beta}\over{b^2}}\  d\beta^2 \nonumber\\ 
&& +\ \left ( 1 \mp {{P^6}\over{r^6_\beta}}\right ) \left ( 2\ dtdr\ +\ {{2r_{\beta}^2}\over{b}}\left ( dt + dr \right )\ d\psi\ +\ {{r^4_\beta}\over{b^2}}\  d\Omega^2_3 \right)\ .\label{dS-2}
\eea
The constants $b$ and $P$ may respectively be identified with a cosmological scale and an energy scale leading to dS geometries in a brane window. 
Across a horizon the light cone flips by a right angle which underlies an interchange of a space-like coordinate with a time-like. Under an interchange $dr\leftrightarrow dt$ across a cosmological horizon, the causal patches may seen to alter the vacuum energy density from a positive to a negative value in a brane window. However the remaining geometric patches in (\ref{dS-2}) remain unchanged on a $D_5$-brane within a vacuum pair. An effective AdS on an anti $D_5$-brane is worked out under the flip of a light-cone. The vacuum geometries on an anti-brane takes a form:
\bea
ds^2&=&\left (1+{{r^2_{\beta}}\over{b^2}}\ \mp\ {{P^6}\over{r^6_\beta}}\right ) dt^2\ +\ 
\left (1 +{{r^2_{\beta}}\over{b^2}}\ \pm\ {{P^6}\over{r^6_\beta}}\right )^{-1} dr^2 \nonumber\\
&&-\ {{2r_{\beta}}^2\over{b}}\left ( dt + dr \right )d\psi\ +\ {{r^2r^2_\beta}\over{b^2}}\  d\beta^2 \nonumber\\ 
&& +\ \left ( 1 \mp {{P^6}\over{r^6_\beta}}\right ) \left ( 2\ dtdr\ +\ {{2r_{\beta}^2}\over{b}}\left ( dt + dr \right )\ d\psi\ +\ {{r^4_\beta}\over{b^2}}\  d\Omega^2_3 \right)\ .\label{dS-2ads}
\eea
An anti-brane moves in an opposite direction to that of a brane within a vacuum created pair $(D{\bar D})_4$-brane at the cosmological horizon. Thus a brane universe is separated from an anti-brane by a null surface and hence their pair annihilation is forbidden in the formalism. A vacuum geometry on an anti brane, within a pair of $(D{\bar D})_4$-brane, may formally be obtained from that on a brane (\ref{dS-2}) under $r\rightarrow -r$. This is supported by the conservation law in a pair creation process. Thus an anti-brane may be obtained from brane under a $\pi$ rotation. 
In fact an observer on a vacuum created brane universe is unaware of the existence of an anti-brane universe. Nevertheless the effective curvature formalism ensures that a brane universe is always associated with an extra hidden dimension which in turn couples to an anti-brane. It is thought provoking to imagine that a higher form (a two form in the case) quanta in a $U(1)$ gauge theory may source the conjectured dark energy in a $(D{\bar D})_3$-brane universe. A schematic scenario of a pair creation is depicted in Figure 3. 
The quantum geometries on a vacuum created $(D{\bar D})_4$-brane is worked out with a real angular momentum ${\tilde J}=(P/b)^6$ at the cosmological horizon.  They are given by
\bea
ds^2&=&-\left (1-{{r^2_{\beta}}\over{b^2}}- {{P^6}\over{r^6_\beta}}\right ) dt^2 +  
\left (1 -{{r^2_{\beta}}\over{b^2}}+ {{P^6}\over{r^6_\beta}}\right )^{-1} dr^2\ +\ {{r^2r^2_\beta}\over{b^2}} d\Omega^2_4\nonumber\\ 
&&-\  {{P^6}\over{b^2r_{\beta}^4}}\left (2b\ d\psi dt + r_{\beta}^2 d\Omega^2_3\right )\ .\label{dS-2a}
\eea
and
\bea
ds^2&=&-\left (1-{{r^2_{\beta}}\over{b^2}}+ {{P^6}\over{r^6_\beta}}\right ) dt^2 +  
\left (1 -{{r^2_{\beta}}\over{b^2}}- {{P^6}\over{r^6_\beta}}\right )^{-1} dr^2\ +\ {{r^2r^2_\beta}\over{b^2}} d\Omega^2_4\nonumber\\ 
&&+\  {{P^6}\over{b^2r_{\beta}^4}}\left (2b\ d\psi dt + r_{\beta}^2 d\Omega^2_3\right )\ .\label{dS-2b}
\eea
\begin{figure}
\centering
\mbox{\includegraphics[width=.5\linewidth,height=0.27\textheight]{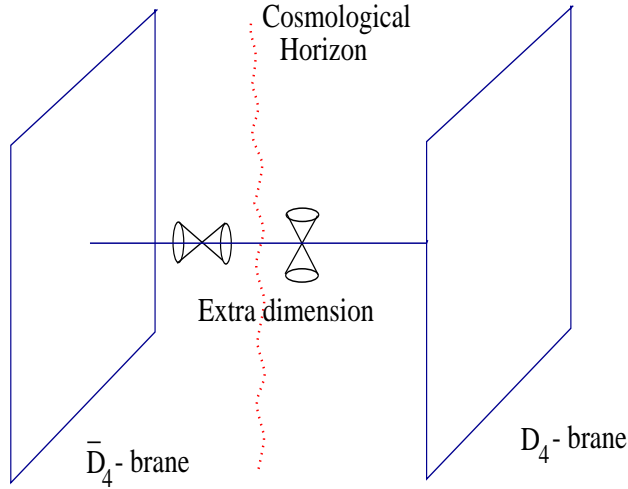}}
\caption{\it Cosmological creation of a $(D{\bar D})_4$-brane pair at a horizon by a non-linear quanta in a two form gauge theory on a $D_5$-brane. An extra sixth dimension is transverse to the five dimensional brane and anti-brane world-volumes.}
\end{figure}

\noindent
A priori, the geometries possess a curvature singularity at $r\rightarrow 0$. Nevertheless the singularity is not accessible in a brane window.
Thus a torsion, $i.e.\ P\neq 0$, may seen to play a significant role to describe a dynamical brane universe in the formalism. 
The cosmological constant in an effective curvature vacuum (\ref{dS-2a}) satisfies ${\Lambda}>0$ at the horizon(s). In absence of a torsion the background metric describes $S^4$ symmetric pure de Sitter vacuum in six dimensions. It is given by
\be
ds^2=-\ \left (1-{{r^2_{\beta}}\over{b^2}}\right ) dt^2\ +\  
\left (1 -{{r^2_{\beta}}\over{b^2}}\right )^{-1} dr^2 \ +\ {{r^2r^2_\beta}\over{b^2}} d\Omega^2_4
\ .\label{dS-3}
\ee
The emergent metric is non-degenerate in a global scenario. For $\beta=\pi/2$, the vacuum geometry precisely identifies with the pure de Sitter underlying a $D_4$-brane in the formalism \cite{spsk-JHEP}. The five dimensional de Sitter vacuum is given by
\be
ds^2=-\ \left (1-{{r^2}\over{b^2}}\right ) dt^2\ +\  \left (1 -{{r^2}\over{b^2}}\right )^{-1} dr^2 \ +\ {{r^4}\over{b^2}}\ d\Omega^2_3\ .\label{dS-4}
\ee
The de Sitter patches in (\ref{dS-3}) and (\ref{dS-4}) are sourced by a global NS two form leading to an open string effective metric on an anti  $D_5$-brane. A higher form (Poincare dual to a gauge field $A_{\mu}$) in the non-linear $U(1)$ gauge theory on an appropriate $D_p$-brane for $p>3$ has been argued to create a vacuum pair of $(D{\bar D})_{p-1}$-brane at the cosmological horizon of a background de Sitter on a $D_p$-brane. Interestingly the cosmological horizon radius in a six dimensional vacuum, for $\beta\neq \pi/2$, is less than that in five dimensions, $i.e.\ r_{\beta}^c<r^c$. It reassures a higher energy in a six dimensional vacuum to that in five dimensional de Sitter brane. The observation is in agreement with a fact that a primordial $4D$ brane universe in the present era has moved a long path away from its cosmological horizon where universe was vacuum created with a Big Bang \cite{spsk-JHEP}. It may also hint at a more fundamental theoretical formulation presumably underlying a geometric 
torsion at a higher dimension. Since a $D_9$-brane is space filling, it provokes thought to believe that a vacuum pair of $(D{\bar D})_9$-brane in an effective curvature formalism may possibly be described in an eleven dimensional $M$-theory. 

\subsection{Quantum correction}
The geometric patches, on a vacuum created pair of $(D{\bar D})_4$-brane, may be rearranged to view as a pure de Sitter phase in presence of a torsion. In particular the pure de Sitter geometric patches are described by an open string metric sourced by a global NS two form on an anti $D_5$-brane. The global $B^{(NS)}_2$ in a gauge invariant combination with an electro-magnetic field is known to describe a non-linear electric charge on a ${\bar D}_5$-brane. Thus the gauge dynamics on a ${\bar D}_5$-brane may be approximated by a Dirac-Born-Infeld (DBI) action. However the open string effective metric serves as a background on a vacuum created $D_5$-brane whose dynamics is in principle governed by a three form (Poincare dual to a gauge field $A_{\mu}$). For simplicity we have explored a dynamical two form on a $D_5$-brane. The two form quanta in a $U(1)$ gauge theory is argued to create a pair of $(D{\bar D})_4$ at a cosmological horizon of a background pure de Sitter on a $D_5$-brane. A pair of brane and anti-
brane universe underlie a six dimensional effective torsion curvature in the formalism. 

\sp
\noindent
Using a brane window we re-express the de Sitter geometric patches (\ref{dS-2}) obtained on a vacuum created pair of $(D{\bar D})_4$-brane to address a dynamical torsion contribution in a second order formalism. The splitted geometric patches, underlying a global NS two form and a local two form, may be given by
\bea
ds^2&=&-\ \left (1-{{r^2_{\beta}}\over{b^2}} \right ) dt^2\ +\ \left ( 1 -{{r^2_{\beta}}\over{b^2}}\right )^{-1} dr^2\ +\ {{r^2r^2_\beta}\over{b^2}} d\Omega^2_4\ \nonumber\\ 
&&\mp\ {{P^6}\over{r^6_\beta}}\left (-dt^2\ +\ dr^2\ +\ {{2r_{\beta}^2}\over{b}}dtd\psi\ +\ {{r^4_{\beta}}\over{b^2}}\ d\Omega_3^2\right )\ .\label{QC-1}
\eea
The open string effective metric defined with an electric non-linear charge $b$ incorporates a positive vacuum energy density. It describes an $S^4$ symmetric de Sitter brane background in a non-linear $U(1)$ gauge theory on a $D_5$-brane. The de Sitter is characterized by a cosmological horizon at $r_{\beta}\rightarrow b$. A conserved charge $P^3$ signifies the presence of a torsion there. It incorporates a (lower dimensional) quantum correction(s) into a $6D$ de Sitter underlying a low energy string vacuum.  

\sp
\noindent
The near cosmological horizon geometries, $i.e.\ r_{\beta}= \left (b\pm \epsilon\right )$, on a vacuum pair of $(D{\bar D})_4$-brane may formally be worked out to yield: 
\bea
ds^2_{\rm NH}&=&-\ \left (1-{{r^2_{\beta}}\over{b^2}} \right ) dt^2\ +\ \left ( 1 -{{r^2_{\beta}}\over{b^2}}\right )^{-1} dr^2\ +\ {{r^2r^2_\beta}\over{b^2}} d\Omega^2_4\ \nonumber\\ 
&&\mp\ {{P^6}\over{r^6_\beta}}\left (-dt^2\ +\ dr^2\ +\ 2b\ dtd\psi\ +\ b^2\ d\Omega_3^2\right )\nonumber\\
&=&-\ \left (1-{{r^2_{\beta}}\over{b^2}} \right ) dt^2\ +\ \left ( 1 -{{r^2_{\beta}}\over{b^2}}\right )^{-1} dr^2\ +\ {{r^2r^2_\beta}\over{b^2}} d\Omega^2_4\nonumber\\
&&\mp\ {{P^6}\over{r^6_\beta}}\left (-dt^2\ +\ dr^2\right )\ .\label{QC-2}
\eea
The off-diagonal metric component signifies a spin angular momentum underlying a torsion in the brane geometry. It shows that the spin and the $S^3$ symmetric patch decouple to confirm a flat metric with the torsion correction. Interestingly a quantum correction is sourced by a dynamical two form which presumably sources a $D$-string. Hence the correction terms may be identified with a non-perturbative contribution underlying $D$-string in a near (cosmological) horizon regime. At this point we digress to recall an analysis discussed in the recent past \cite{spsk-JHEP}. The pair production mechanism by a two form gauge theory on a $D_4$-brane has been argued to vacuum create a pair of $(D{\bar D})$-instanton which grows to describe a pair of of $(D{\bar D})$-particle followed by the higher dimensional vacuum pairs: $(D{\bar D})$-string, $(D{\bar D})$-membrane and $(D{\bar D})_3$-brane. 

\sp
\noindent
Analysis reveals a nullifying effect of angular momentum on a vacuum created $D_4$-brane by that on an anti $D_4$-brane, $i.e.$ under $b\rightarrow -b$. 
The quantum effects are incorporated into a low energy string vacuum by the world-volume of a $D_4$-brane. The non-perturbative ($NP$) quantum correction may be given by
\be
ds^2_{\rm NP}\ =\ -\ dt^2\ +\ dr^2 \ +\ {{r^4}\over{b^2}}\ d\Omega_3^2\ .\label{QC-3}
\ee
The Ricci curvature scalar is computed for the associated geometry on a vacuum created $D_4$-brane to yield:
\be
R= \left ( {{b^2}\over{r^4\sin^2\psi}} + {{3b^2}\over{r^4}} - {{35}\over{r^2}}\right )\ .\label{QC-4}
\ee
The brane window $P<r<b$ does not access the curvature singularity which is otherwise prevailed in the emergent geometric correction. It may be worked out for $\psi=(\pi/4)$ at the cosmological horizon. Then the curvature scalar becomes
\be
R_{\rm r\rightarrow b}=\ -\ {{30}\over{b^2}}\ .\label{QC-5}
\ee
The $(3+1)$-dimensional geometric correction presumably underlie a vacuum created $D_3$-brane. The brane geometry identifies with an AdS curvature scalar $R$ and hint at two extra dimensions in the formalism. A negative energy (density) correction via a $D_3$-brane to a stringy de Sitter is noteworthy in the case. Interestingly the Ricci curvature scalar computed in presence of a spin angular momentum for the metric in a torsion correction (\ref{QC-1}) precisely identifies with (\ref{QC-4}). It re-confirms our assertion that a spin angular momentum is nullified in the emergent geometries on a vacuum created pair of $(D{\bar D})_4$-brane.  The curvature scalar (\ref{QC-4}), for $\psi=(\pi/4)$, vanishes at a fixed point $r\rightarrow r_0=\left (b/{\sqrt{7}}\right )$ which turns out to be in the near horizon regime. A flat metric coupling to a torsion charge $P^3$ at a fixed point $r_0$ is consistently described in a brane window $P<r<b$. It reassures a lower dimensional $D$-brane, $i.e.$ a $D_3$-brane, world-
volume correction to a low energy closed string vacuum.  In fact a string vacuum is indeed sourced by a global NS two form on an anti $D$-brane. It may also be sensed by an extra sixth transverse dimension between a vacuum created pair of $(D{\bar D})_4$-brane underlying an effective torsion curvature. It is interesting to note that a perturbative (low energy) closed string vacuum in a type II superstring theory receives a non-perturbative quantum correction underlying a (lower dimensional) vacuum created $D$-brane or an anti $D$-brane world-volume in the formalism. 

\sp
\noindent
The quantum ($D$-brane world-volume) correction in (\ref{dS-2}) may be re-expressed using the light-cone coordinates: $x_{\pm}=(t\pm r)$. The fluctuations on a vacuum created brane universe is given by
\be
ds^2_{q+}=\ \mp\ {{P^6}\over{b^2r^4_\beta}}\left ( {{b^2}\over{r^2_{\beta}}} dx_+^2\ +\ 2b\ dx_+d\psi\ +\ r^2_{\beta}\ d\Omega_3^2\right )\ .\label{QC-6}
\ee
Similarly a correction by an anti $D_4$-brane may be given by
\be
ds^2_{q-}=\ \mp\ {{P^6}\over{b^2r^4_\beta}}\left ( {{b^2}\over{r^2_{\beta}}} dx_-^2\ +\ 2b\ dx_-d\psi\ +\ r^2_{\beta}\ d\Omega_3^2\right )\ .\label{QC-7}
\ee
The quantum fluctuations leading to the emergent geometries on a vacuum created $D_5$-brane and an anti $D_5$-brane are manifestations of a geometric torsion ($P\neq 0$). They are independently described on a brane and an anti-brane respectively by the radial coordinates $x_+$ and $x_-$. Unlike to a vacuum created $D_4$-brane correction in (\ref{QC-1}), the light-cone coordinates assure one lower dimensional brane, $i.e.$ a $D_3$-brane, correction to the low energy string vacuum. It is due to a fact that a world-volume dimension appears to decouple in a light-cone coordinates. Furthermore, the line-element, sourced purely by a torsion fluctuations, reduces to describe a vacuum created $D_2$-brane. It may be re-expressed as:
\bea
&&ds^2_{q\pm}\ =\ \mp\ {{P^6}\over{r^6_\beta}}\left ( d\rho_{\pm}^2 \ +\ {{r^4_{\beta}}\over{b^2}} \sin^2\psi\ d\Omega_\theta^2\right )\ ,\nonumber\\
{\rm where}&&\quad d\rho_{\pm}=\ \left ( dx_{\pm}\ +\ {{r_{\beta}^2}\over{b}} d\psi\right )\ .\label{QC-8}
\eea
At the creation of a vacuum pair of $(D{\bar D})_2$-brane, $i.e.$ in a limit to a cosmological horizon $r_c\rightarrow b(\sin\beta)^{-1}$, the new radial coordinates $\rho_{\pm}$ may be expressed in terms of the original coordinates $(t,r,\psi)$. It may formally be expressed as: $\rho_{\pm}^c= (t +r+ b\psi)_c= (x_{\pm}+b\psi)_c$. The near (cosmological) horizon line-element leading to a non-perturbative correction simplifies to yield:
\be
ds^2_{q\pm} \rightarrow \left ( d\rho_{\pm}^2\ +\ b^2 \sin^2\psi\ d\Omega_\theta^2\right )\ .\label{QC-9}
\ee
It reassures that an $S^2$ geometry decouples from a vacuum created $D_2$-brane in a near (cosmological) horizon geometry which in turn describes a $D_0$-brane. It further shows that the near horizon causal patches become flat. Arguably a low energy perturbative string vacuum may seen to receive 
a non-perturbative correction underlying a vacuum created pair of $(D{\bar D})$-instanton in an effective curvature formalism. The world-volume dimensions is believed to unfold through the quantum corrections. Intuitively the geometric evolutions of space-time began with a vacuum created pair 
of $(D{\bar D})$-instanton with a Big Bang at a cosmological horizon. A growth in the space-time dimension may be viewed through a series of nucleation of a higher dimensional brane pair from a lower brane. The nucleation ceases with a vacuum pair of $(D{\bar D})_4$-brane underlying a non-linear $U(1)$ gauge theory on a $D_5$-brane.

\sp
\noindent
The emergent de Sitter geometries (\ref{dS-3}) on a pair of $(D{\bar D})_4$-brane is analyzed at the cosmological horizon using Painleve transformations. It may be given by
\be
ds^2=-\ \left (1-{{r^2_{\beta}}\over{b^2}}\right ) dt^2\ +\ dr^2\ \pm\  {{2r_{\beta}}\over{b}} dt dr\ +\ {{r^2r^2_\beta}\over{b^2}} d\Omega^2_4
\ .\label{QC-10}
\ee
In the limit $r_{\beta}\rightarrow r_c=b$ the effective de Sitter metric reduces to yield:
\be
ds^2\rightarrow \ \left ( dr^2 \ \pm\ 2dt dr\ +\ r^2 d\Omega^2_4\right )\ .\label{QC-11}
\ee
Interestingly the emergent geometries may be identified with a six dimensional pure de Sitter in Painleve coordinates established in Einstein vacuum.
The torsion dynamics in the formalism has been shown to describe a quintessence axion on an anti $D_3$-brane universe within a vacuum created pair \cite{psskk,pssk1,pssk2}. A quintessence is believed to be a potential candidate to source the conjectured dark energy in our universe.

\section{Quantum de Sitter in ${\mathbf{6D}}$: Positive and negative mass pairs}
\subsection{Discrete transformation within causal patches}
The effective de Sitter vacua (\ref{dS-2a}) neither correspond to a Schwarzschild de Sitter (SdS) nor to a Topological de Sitter (TdS) causal patch known in Einstein vacuum. However they may be viewed as an orthogonal combination of a Schwarzschild and a topological metric components: $G_{tt}$ and $G_{rr}$.
We perform a matrix projection \cite{spsk-JHEP} to project out the SdS and TdS causal patches from a mixed phase in the quantum regime. In the case we define a ($2\times2$) matrix $N$ containing the longitudinal components of the metric quanta with euclidean signature. It is given by
\begin{equation}
N=\frac{1}{2}\left( \begin{array}{ccc}
G_{tt}(-)&&G_{rr}(-)\\
&& \\
G_{rr}(+)&&G_{tt}(+)
\end{array} \right)\ ,\label{SdS-1}
\end{equation}
where
\bea
&&G_{tt}(-)= \left (1-{{r^2_{\beta}}\over{b^2}} - {{P^6}\over{r^6_\beta}}\right )\ ,\;\; G_{rr}(-)= \left (1-{{r^2_{\beta}}\over{b^2}} - {{P^6}\over{r^6_\beta}}\right )^{-1}\ ,\nonumber\\
&&G_{tt}(+)= \left (1-{{r^2_{\beta}}\over{b^2}} + {{P^6}\over{r^6_\beta}}\right )\;\; {\rm and}\;\;
G_{rr}(+)= \left (1-{{r^2_{\beta}}\over{b^2}} + {{P^6}\over{r^6_\beta}}\right )^{-1}\ .\label{SdS-2}
\eea
The metric components in $G_{\mu\nu}(-)$ and $G_{\mu\nu}(+)$ are respectively associated with SdS 
and TdS geometries. A matrix projection, on the column vectors, yields:
\begin{equation}
N\left( \begin{array}{c}1\\
\\
0
\end{array}\right)=\frac{1}{2}\left( \begin{array}{c}
G_{tt}(-)\\
\\
G_{rr}(+)
\end{array}\right) 
\qquad{\rm and} \qquad N\left( \begin{array}{c}
0\\
\\
1
\end{array}\right)=\frac{1}{2}\left( \begin{array}{c}
G_{rr}(-)\\
\\
G_{tt}(+)
\end{array}\right)\ .\label{SdS-3}
\end{equation}
The projection geometries indeed describe a mixed causal patch obtained in eq(\ref{dS-2a}). The ($\det N$) may be computed to yield
\be
\det N= -{{r^2_{\beta}}\over{b^2}}\ .\label{SdS-4}
\ee
The determinant at the cosmological horizon ensures ($\det N=-1$) a discrete transformation. The matrix inverse is computed to yield:
\begin{equation}
N^{-1}=\frac{1}{2\det N}\left( \begin{array}{ccc}
G_{tt}(+) && -G_{rr}(-)\\
&& \\
-G_{rr}(+)&&G_{tt}(-)
\end{array} \right)\ . \label{SdS-5}
\end{equation}
Interestingly an inverse matrix projection on the same column vectors separate out the SdS potential from the TdS. They are given by
\begin{equation}
 N^{-1}\left( \begin{array}{c}
1\\
\\
0
\end{array}\right)=\frac{1}{2}\left( \begin{array}{c}
-G_{tt}(+)\\
\\
G_{rr}(+)
\end{array}\right) 
\;\; {\rm and} \;\; N^{-1}\left( \begin{array}{c}
0\\
\\
1
\end{array}\right)=\frac{1}{2}\left( \begin{array}{c}
G_{rr}(-)\\
\\
-G_{tt}(-)
\end{array}\right) \ . \label{SdS-6}
\end{equation}
It shows that the effective longitudinal metric components (\ref{dS-2a}) under a discrete transformation reestablish the lorentzian signature.

\subsection{Topological de Sitter brane}
The inverse matrix operation on a (spin up) column vector in (\ref{SdS-6}) projects out a topological de Sitter (TdS) causal patches in an effective curvature formalism. The microscopic TdS black hole in $6D$ is indeed sourced by a $D_5$-brane world-volume torsion charge $p$ $(=P^3)$ in a $U(1)$ gauge theory. Then an effective TdS on a vacuum pair of $(D{\bar D})_4$-brane, underlying a $D_5$-brane, is given by
\bea
ds^2&=&-\left (1-{{r^2_{\beta}}\over{b^2}} + {{P^6}\over{r^6_\beta}}\right ) dt^2 +  
\left (1 -{{r^2_{\beta}}\over{b^2}} + {{P^6}\over{r^6_\beta}}\right )^{-1} dr^2\  +\ {{r^2r^2_\beta}\over{b^2}} d\Omega^2_4\nonumber\\
&&+\  {{P^6}\over{b^2r_{\beta}^4}}\left (2b\ d\psi dt + r_{\beta}^2 d\Omega^2_3\right )\ .\label{SdS-8}
\eea
In absence of a dynamical torsion, $i.e.$ for $p=0$, the emergent geometry reduces to a $6D$ de Sitter in a low energy string vacuum. It is given by
\be
ds^2=-\left (1-{{r^2_{\beta}}\over{b^2}}\right ) dt^2 +  \left (1 -{{r^2_{\beta}}\over{b^2}}\right )^{-1} dr^2\  +\ {{r^2r^2_\beta}\over{b^2}} d\Omega^2_4\ .\label{SdS-81}
\ee
An electric string charge $+b$ may be seen to describe a cosmological horizon on a brane universe. Then, an anti-brane would like to be described by $-r$ and $-b$ with respect to a brane, though a brane or an anti-brane universe are independently characterized by $(r,b)$. 

\sp
\noindent
On the other hand for $p\neq 0$, an extremely large string charge $b$ may seen to describe an isolated topological quantum gravity without a classical analogue as $P^6>0$. This is due to a fact that the quantum geometry forbids a point notion underlying a smeared (non-fundamental string) charge $P^3$ or mass ($P^6$) and hence $r\rightarrow 0$ is forbidden. Nevertheless the smearing disappears in a classical limit which in turn 
re-establishes a point notion allowing $r\rightarrow 0$. Thus a classical limit would describe a naked singularity and hence the emergent geometry would not describe a physical brane universe. The topological quantum geometry on a vacuum pair of $(D{\bar D})_4$-brane is given by
\be
ds^2=-\left (1+ {{M}\over{r^6_\beta}}\right ) dt^2 +  \left (1 + {{M}\over{r^6_\beta}}\right )^{-1} dr^2\ +\ r_{\beta}^2 d\Omega^2_3\ ,\label{SdS-82}
\ee
where the square of non-linear charge may formally be identified with a gravitational mass $M=p^2$. It implies $M>0$ in a TdS black hole (\ref{SdS-8}).
Needless to mention that a negative mass in the geometry would incorporate an event horizon which in turn drains off the topological notion.
An observer in a classical TdS black hole is protected from a curvature singularity by a (cosmological) horizon at $r\rightarrow {\tilde r}_{c}=\left (b\sin^{-1}\beta-\delta M\right )$. However the horizon in a quantum TdS is not accessible in the brane window $M<r<b$. This is in conformity with the notion of a smeared mass set by an ultra-violet cut-off in the gauge theory. Numerical analysis further reassures the assertion. 

\subsection{Schwarzschild de Sitter brane}
Under an inverse matrix operation on a (spin down) column vector in (\ref{SdS-6}) we obtained the required causal patches for an effective Schwarzschild de Sitter brane. The microscopic SdS black hole in $6D$ is indeed sourced by a $D_5$-brane world-volume two form quanta $p$ in a $U(1)$ gauge theory. The SdS black hole on a vacuum pair of $(D{\bar D})_4$-brane underlying a $D_5$-brane is given by
\bea
ds^2&=&-\left (1-{{r^2_{\beta}}\over{b^2}}-{{M}\over{r^6_\beta}}\right ) dt^2 +  
\left (1 -{{r^2_{\beta}}\over{b^2}}- {{M}\over{r^6_\beta}}\right )^{-1} dr^2\  +\ {{r^2r^2_\beta}\over{b^2}} d\Omega^2_4\nonumber\\
&&-\ {{M}\over{b^2r_{\beta}^4}}\left (2b\ d\psi dt + r_{\beta}^2 d\Omega^2_3\right )\ .\label{SdS-7}
\eea
The six dimensional quantum SdS black hole is characterized by a cosmological horizon at $r\rightarrow r_c=(b\sin^{-1}\beta-\delta M)$ and an event horizon at $r\rightarrow r_e=(M\sin^{-1}\beta +\delta M)$. Interestingly the horizons are accessible in the brane window $M<r<b$. Numerical analysis for various set values of $(M,b)$ is in agreement with the qualitative result. 

\sp
\noindent
Furthermore a negative mass $M$ in SdS brane becomes physical as the curvature singularity at $r\rightarrow 0$ is not accessible in a brane window. A minimal length scale in $r$ re-ensures a non-linear electric or magnetic charge. Interestingly the notion of a negative mass in SdS brane 
is analogous to the idea \cite{paranjape} of a negative mass in SdS black hole coupled to a perfect fluid. Thus $M<0$ in SdS brane and $M>0$ in TdS brane may be viewed via an electro-magnetic duality on a $D_3$-brane. Formally an electric non-linear charge $p_e=p$ in SdS brane may be replaced with a non-linear magnetic charge $p_m=ip_e$ to yield a TdS brane. Then the quantum geometry in eq(\ref{SdS-82}) may be viewed from a pure de Sitter vacuum under a flip of light-cone across the cosmological horizon. It may imply that an emergent pair of brane universe, created by a (non-fundamental) 
stringy quanta at a dS (perturbative) string vacuum, describes $M<0$ in a brane and $M>0$ in an anti-brane or vice-versa. Thus the vacuum pair across a cosmological horizon may be identified with a SdS brane universe and a TdS anti-brane universe. Generically a negative mass black hole is known to describe a repulsive gravity (anti gravity). A negative gravitational mass has been argued to produce in pair with a positive mass. A negative gravitational mass repels a positive mass while a positive gravitational mass attracts a negative mass. Thus a negative mass and a positive mass is believed to move in a pair.

\sp
\noindent
A priori the SdS geometry possesses a curvature singularity at $r\rightarrow 0$. However a lower bound on $r$ prescribed by a brane window rules out the possibility of a curvature singularity in the quantum SdS black hole (\ref{SdS-7}). The absence of a curvature singularity in a black hole may also be seen to be influenced by the quantum effects incorporated by the gauge theoretic quanta on a $D$-brane into the background string vacuum. An electric non-linear charge on a $D$-brane endorses a minimal non-zero length scale which in turn incorporates an ultra-violet cut-off. Hence a non-linear charge endorses a string scale on a brane which does not allow $r\rightarrow 0$. Hence the obtained quantum SdS brane geometry is free from the curvature singularity and may be viewed as a trapped regime. 

\sp
\noindent
Analysis reveals that an observer in SdS brane universe is restricted to a time-like regime between the cosmological horizon $r_c$ and an event horizon $r_e$. The horizons are essentially sourced by two independent  potentials underlying two distinct electric non-linear charges: a fundamental string charge $b$ and a (non-fundamental) torsion charge $p$. An instability arised out of the potential difference between the two horizons has been argued to intiate Hawking radiation phenomenon in the SdS black hole. As a result, the event horizon expands to $r_e\rightarrow {\hat r}_e=(M\sin^{-1}\beta +n\delta M)$ and the cosmological horizon shrinks to $r_c\rightarrow {\hat r}_c=(b\sin^{-1}\beta-m\delta M)$, where ($n,m$) are arbitrary integers. The expansion of an event horizon may be interpreted due to a growth in $M$. It may also be viewed through an effective geometric torsion curvature underlying a growth in dark energy in a brane universe. 

\sp
\noindent
The SdS brane has been argued to evaporate to a Nariai brane geometry \cite{spsk-JHEP}. Though the horizon areas are equal in the Nariai limit, they
are separated along a time-like coordinate with a non vanishing phase difference, $(\delta M)_t\rightarrow |r_c-r_e|_t$, between the space-like equipotential. In fact, the interpolating potential between the two horizons may be worked out to yield a global maximum in the geometric phase defined by a temporal $r$ in the regime.  In particular the equipotential with a spatial $r$ may be viewed at 
$r_e\rightarrow r_c$. A non-zero geometric time-like phase $(\delta M)_t$ may be analyzed between the two horizons at $r_e$ and $r_c$. It shows that the potential starts to increase from $(M\sin^{-1}\beta+n\delta M)=M_{max}$ to a maximum and then falls to arrive at an equipotential at $(b\sin^{-1}\beta-m\delta M)=b_{min}$. Intuitively, the maximum along a temporal $r$ may be interpreted as a shock wave peak along a spatial $r$ coordinate at $r\rightarrow r_e\rightarrow r_c$ in the SdS brane. The expansion of event horizon ceases when the equipotential is approached. In the limit, the mass becomes maximum and takes a critical value $M_{max}$. The maximal mass is formally identified with the Nariai mass in a de Sitter black hole. It may imply that the Hawking radiations underlying the instabilities transform the SdS brane into a Nariai black hole in the regime. The TdS black hole has been shown to describe a meta-stable phase of the Nariai black hole in the quantum geometry \cite{cai-myung-zhang,rong-cai}.

\subsection{Spin angular momentum}
The angular velocities in the SdS brane and in the TdS brane are computed at their cosmological horizons. They are given by
\be
\Omega_{r_c}^\psi({\rm SdS})=\ -\ \Omega_{r_c}^\psi({\rm TdS})=\ -\ \frac{M}{b^7}\ .\label{SdS-9}
\ee
It is evident that the effective (SdS and TdS) black holes are rotating with a small angular velocity in opposite directions to each other at their cosmological horizons. The angular velocity of SdS at the event horizon is computed to yield
\be
\Omega_{r_e}({\rm SdS})=\ J{{b}\over{P^2}}\ .\label{SdS-10}
\ee
Thus a SdS brane is rotating with a very large angular velocity at its event horizon $r_e$. We recall that a pure dS at the Big Bang was primarily sourced by a coupling of a global NS two form in disguise of a non-propagating torsion in an effective curvature formalism. The dynamical effects of a geometric torsion began at the cosmological horizon with the creation of a vacuum pair of $(D{\bar D})_4$ underlying a positive-negative mass pair. A slow rotation at a cosmological horizon in brane and anti-brane universes reassure a small torsion or mass at the Big Bang. An increase in angular velocity at the event horizon in a SdS brane or an anti-brane universe reflects an enhanced strength in torsion and hence a large mass there.

\sp
\noindent
On the other hand the SdS and TdS brane universes differ significantly in their potential from the de Sitter vacuum in Einstein gravity  \cite{cai-myung-zhang,rong-cai,medved}. For instance the effective potentials in the six dimensional effective curvature formalism hint at a nine dimensional space-time at Planck scale. It confirms the presence of three extra transverse dimensions hidden to a brane universe. The extra dimensions between a pair of $(D{\bar D})_4$-brane may be viewed through the ten dimensional type IIA and type IIB superstrings on $S^1$. Furthermore the extra dimensions may hint at a vacuum created pair of $(D{\bar D})_8$-brane universe at Planck scale in an effective curvature formalism. A higher dimensional vacuum pair is in conformity with a space filling $D_9$-brane in a ten dimensional type IIB superstring theory and their pair, $i.e.\ (D{\bar D})_9$-brane, may be described in an eleven dimensional M-theory.

\subsection{Thermal analysis}
A global NS two form in addition to a dynamical two form do play a significant role to define the vacuum energy underlying a $D_5$-brane. 
In particular, a global NS two form leads to an open string effective metric and hence sources a non-zero potential. A dynamical two form incorporates the quantum fluctuations into a background pure de Sitter in a $U(1)$ gauge theory on a $D_5$-brane. The local degrees do play a significant role to compute the total vacuum energy in a de Sitter brane universe. It may qualitatively be analyzed from a generalized notion of an emergent metric for its $G_{tt}$ component. The total energy function is worked out to yield: 
\be
E(r)= \left ( G_{tt}-g_{tt} \right )\ .\qquad {}\label{T-1} 
\ee
The energy function in an effective (SdS and TdS) black holes with a lorentzian signature are respectively given by
\be
E_S^l(r)\ =\ \left ( {{r^2_{\beta}}\over{b^2}} + {{M}\over{r^6_\beta}}\right )\quad 
{\rm and}\qquad E_T^l(r)\ =\ \left ( {{r^2_{\beta}}\over{b^2}} - {{M}\over{r^6_\beta}}\right )\ .\label{T-2}
\ee
The energy density in a SdS brane black hole possesses a minimum at $r_0= ({\sqrt{3}}\ r_cr_e^3)^{1/4}$. The vacuum energy at its spatial horizon 
$r_c=\left (b_{min}\sin^{-1}\beta\right )=r_e=\left (P_{max}\sin^{-1}\beta\right )$ is estimated to yield:
\bea
E_N^l(r)|_{r_e}&=&\left ( 1 + {{P^2_{max}}\over{b^2}}\right )_P\nonumber\\
&\rightarrow& \left ( 1 + {{M_{max}}\over{b^6}}\right )_b\approx 1\ .\label{T-4}
\eea 
The vacuum energy in the TdS brane at a cosmological horizon $r_c=\left (b\sin^{-1}\beta\right )-\delta M$ becomes
\bea
E_T^l(r)|_{r_c}&=&\left ( 1 - {{M}\over{b^6}}\right )\ +\ {{2\delta M\sin\beta}\over{b}}\left (1 + {{3M}\over{b^6}}\right )\nonumber\\
&\rightarrow&\left ( 1 +\ {{2\delta M\sin\beta}\over{b}}\right ) \approx 1\ .\label{T-5}
\eea 
It shows that the energy in a Nariai vacuum equals to that in the TdS brane and they are defined at the same spatial horizon: $M_{max}=b_{min}$. However they are separated by a temporal phase: $(\delta M)_t\neq 0$. A maximal torsion in Nariai brane undergoes quantum tunnelling and may be
identified with the TdS brane \cite{parikh-wilczek,cai-myung-zhang,medved,rong-cai}. The maximal energy is used to change a negative gravitational mass in Nariai brane to a positive mass in the TdS brane. The maximal torsion has been identified with a condensate of discrete torsion in TdS brane \cite{spsk-JHEP}. The SdS and TdS black holes may be believed to describe the near (cosmological) horizon $D_5$-brane geometries. Their horizon radii: $\left (b\sin^{-1}\beta\mp\delta M\right )$ further reassure the near horizon brane geometries. An observer in SdS brane would likely to encounter the phenomenon of Hawking radiations from both the horizons. Thermal analysis ensures the direction of net flow towards the cosmological horizon. The gravitational energy in SdS brane increases via Hawking radiations. As a result the SdS brane evaporates and leaves behind a Nariai black hole. From the perspective of an observer in TdS brane, the negative energy (particle) from SdS tunnels to TdS through their near (
cosmological) horizon boundaries. The positive energy anti-particle from a created pair inside the cosmological horizon raises the gravitational energy in the SdS brane. A pair production process, just inside a cosmological horizon, in the TdS brane continues until the (dark) energy attains its maximum $M_{max}$ in the SdS brane. In the limit the cosmological scale reduces to a minimal value $b_{min}$ and may be identified with a condensate.

\sp
\noindent
Furthermore a thermal analysis may allow one to express the temperature in the SdS brane at its event horizon: $({\sqrt{M}}\ \sin^{-1}\beta+\delta M)$. It is given by
\bea
T^{\rm SdS}_{r_e}&=&{1\over{4\pi}} \partial_rG_{tt}{\Big |}_{r_e}\nonumber\\ 
&=&{1\over{2\pi r_e}}\left ( 3-{{4r_e^2\sin^2\beta}\over{b^2}} \right )\nonumber\\
&=&nT_0\left ( 3\sin\beta\,\left (1 -{{\delta M\sin\beta}\over{\sqrt{M}}}\right ) - 4\left ( 1 + {{\delta M\sin\beta}\over{\sqrt{M}}} \right ) {{M\sin\beta}\over{b^2}}\right )\nonumber\\
&\rightarrow&3nT_0\sin\beta\left ( 1 -{{\delta M\sin\beta}\over{\sqrt{M}}}\right )\nonumber\\
&\approx& mT_0\sin\beta\ .\label{qads-21}
\eea
The time-like phase, $\delta M_t\neq 0$ in SdS brane, does not seem to validate the thermal field theoretic technique to compute temperature in SdS at its cosmological horizon $r_c$. Nevertheless, under a change in signature, the SdS vacuum may be identified with the TdS. The notional change in metric signature may provoke thought to re-express the temperature at $r_c$ in a generic dS using a lorentzian metric without a mod in its formal definition. Generically a mod ensures a positive temperature at the horizon. It may be relaxed at the expense of a lorentzian metric at $r_c$. The notion may further be supported by an observation that the quantum geometries may be approximated by a hot fluid at $r_c$. Presumably the notion of time becomes significant with a $D_4$-brane and an anti $D_4$-brane under a generalized description of branes within a higher dimensional brane. Intuitively the temperature in SdS brane at the cosmological horizon may equivalently be given by
\bea
T^{\rm SdS}_{r_c}&=&-{1\over{4\pi}} \partial_r G_{tt}{\Big |}_{r_c}\nonumber\\
&=&{1\over{2\pi r_c}}\left ({{4r_c^2\sin^2\beta}\over{b^2}} -3\right )\ .\label{qads-22}
\eea
The temperature expression holds good for its evaluation at the cosmological horizon. The non-zero time-like phase in the pure dS and in the SdS brane are primarily responsible for the renewed definition. With a subtlety, it may imply that the dS vacua are generically associated with a high temperature presumably with a real time and may enhance our understanding of dual Euclidean CFT. We compute the temperature in SdS brane at cosmological horizon: ($b\sin^{-1}\beta-\delta M)$. It is given by
\bea
T^{\rm SdS}_{r_c}&=&nT_0\left ( 1 -{{7\delta M\sin\beta}\over{b}}\right )\left ({{{\sqrt{M}}\sin\beta}\over{b}}\right )\nonumber\\
&\approx& nT_0 \left ({{\sqrt{M}}\over{b}}\right )\sin \beta\ .\label{qads-23}
\eea
It is evident that the temperature in an emergent SdS brane at $r_c$ is suppressed by a scale factor $({\sqrt{M}}b^{-1})<1$, than its temperature estimated at $r_e$. It re-assures a net flow of Hawking radiations towards $r_c$ in the SdS brane. As a result, the vacuum would eventually evolve 
towards a pure dS. The phenomenon is in agreement with the second law of thermodynamics underlying a fact that the total entropy in an asymptotic SdS is bounded from above by the entropy of a pure dS.

\section{Concluding remarks}
An effective space-time curvature in six dimensions was formalized underlying a two form CFT in presence of a background de Sitter on a $D_5$-brane. In fact the pure de Sitter was obtained with a gauge ansatz for a global NS two form presumably on a (background) anti $D_5$-brane. The set up of a brane dynamics in presence of a static anti-brane may be lifted to a higher dimensional scenario underlying a pair of non-coincident brane and an anti-brane  in presence of an extra seventh dimension transverse to both the world-volumes. 

\sp
\noindent
It was argued that a two form quanta, in the world-volume CFT of a $D_5$-brane, produces a pair of $(D{\bar D})_4$-brane at a cosmological horizon of a pure dS black hole in the background. The non-linearity in the world-volume torsion charge was argued to describe a pair of (non-fundamental) strings. A pluasible string/five brane duality may be invoked to 
describe a $D_5$-brane universe in the formalism. Closed string exchange 
between a pair was intuitively realized with the emerging quantum torsion geometries. The emergent scenario has been shown to describe an extra dimension transverse to a brane (or anti-brane) universe. The hidden dimension to the world-volume was argued to play a significant role 
and may explain the presence of a quintessence scalar in the formalism. In a low energy limit the world-volume torsion becomes weak and hence decouples to describe Einstein gravity. We believe a geometric torsion may be relevant to the study of inflationary cosmology. 
It may provide a clue to unfold a plausible source for dark energy in our universe. 

\sp
\noindent
It may be interesting to generalize the formalism further to describe a higher dimensional brane universe. The construction would involve a higher form on a $D_p$-brane which in turn would like to be described by the M-theory in eleven dimensions. Interestingly the closed string modes in a space filling $D_9$-brane in type IIB superstring theory would be described by a vacuum created pair of $(D{\bar D})_8$-brane. 
Interestingly a pair of $(D{\bar D})_9$-brane shall be annihilated due to the absence of an extra transverse dimension between the world-volume  
in a ten dimensional type IIB superstring theory. It re-assures that a vacuum pair formation stops with a creation of 
$(D{\bar D})_8$-brane. Intuitively an extra dimension between a pair of $(p{\bar p})$-brane for $p=9$ may hint for an eleven dimensional 
non-perturbative construction. Thus an appropriate higher form dynamics on a space filling brane in presence of a background anti $9$-brane may provide a clue to $M$-theory.

\sp
\noindent
We have analyzed the emergent quantum patches on a $(D{\bar D})_4$-brane underlying a geometric torsion in the formalism. The vacuum created 
SdS brane and TdS brane have been analyzed to describe an emergent pair of negative and positive masses at a cosmological horizon. In fact, the dS 
brane geometries were found to be mixed with Schwarschild and topological causal patches. They were separated out with a discrete transformation underlying a matrix projection. Interestingly the quantum dS branes were shown to possess a spin angular momentum which is intrinsic to a torsion in the formalism. Computation of angular momentum in SdS confirms a higher spin at an event horizon than at a cosmological horizon. A thermal analysis was invoked in SdS brane to reveal an equilibrium between the horizons which in turn describes a stable SdS brane.

\sp
\noindent
Interestingly the microscopic SdS brane and TdS brane (or anti-brane) have been shown to be free from a curvature singularity. They have been identified with the low energy perturbative string vacuum in $6D$ in presence of a non-perturbative lower dimensional $D_p$-brane correction for $p<5$. A priori, the quantum correction was shown to evolve with a non-trivial curvature. Analysis reveals that a non-perturbative curvature turns out to be constant at the creation of a pair of $(D{\bar D})_4$-brane universes at a Big Bang singularity.

\section*{Acknowledgments}
We gratefully thank D.S. Kulshreshtha, V. Ravishankar and S. Sarkar for discussions during a presentation of this work by R.K. The research work of R.K. is partly supported by C.S.I.R. New Delhi and the research of D.S. is partly supported by U.G.C. New Delhi. S.K. acknowledges the R\&D research grant 2014 from the University of Delhi.

\def\anp{Ann. of Phys.}
\def\cmp{Comm.Math.Phys.}
\def\prl{Phys.Rev.Lett.}
\def\prd#1{{Phys.Rev.} {\bf D#1}}
\def\jhep{JHEP\ {}}{}
\def\jaat{J.Astrophys.Aerosp.Technol.\ {}} {}
\def\cqg#1{{Class.\& Quant.Grav.}}
\def\plb#1{{Phys. Lett.} {\bf B#1}}
\def\npb#1{{Nucl. Phys.} {\bf B#1}}
\def\mpl#1{{Mod. Phys. Lett} {\bf A#1}}
\def\ijmpa#1{{Int.J.Mod.Phys.}{\bf A#1}}
\def\mpla#1{{Mod.Phys.Lett.}{\bf A#1}}
\def\rmp#1{{Rev. Mod. Phys.} {\bf 68#1}}
\def\ptep{Prog. of Theo.\& Exper.Phys.}
\def\jcap{J.Cosmo.\& Astropar.Phys.}


\begin{thebibliography}{99}

\baselineskip= 18 truept

\bibitem{riess}A.G. Riess et al. [{\it Supernova search team collaboration}], Astron.J.{\bf 116} (1998) 1009.

\bibitem{bousso-hawking}R. Bousso and S. Hawking, \prd {\bf 57} (1998) 2436.

\bibitem{parikh-wilczek}M.K. Parikh and F. Wilczek, \prl {\bf 85} (2000) 5042;

M.K. Parikh, \plb {\bf 546} (2002) 189

\bibitem{hawking-maldacena-strominger}S. Hawking, J.M. Maldacena and A. Strominger, \jhep {\bf 05} (2001) 001

\bibitem{strominger-ds}A. Strominger, \jhep {\bf 10} (2001) 034

\bibitem{cai-myung-zhang}R-G. Cai, Y.S. Myung and Y-Z. Zhang, \prd {\bf 65} (2002) 084019

\bibitem{dehghani}M.H. Dehghani, \prd {\bf 65} (2002) 104030

\bibitem{bousso-maloney-strominger}R. Bousso, A. Maloney and A. Strominger, \prd {\bf 65} (2002) 104039

\bibitem{medved}A.J.M. Medved, \prd {\bf 66} (2002) 124009

\bibitem{klemm}D. Klemm,  \npb {\bf 625} (2002) 295

\bibitem{rong-cai}R-G. Cai, \plb {\bf 525} (2002) 331; R-G. Cai, \npb {\bf 628} (2002) 375

\bibitem{banks-fiol-morisse}T. Banks, B. Fiol and A. Morisse, \jhep {\bf 12} (2006) 004

\bibitem{polyakov}A.M. Polyakov, \npb {\bf 797} (2008) 199

\bibitem{lee-lee}B-H. Lee and W. Lee, \cqg {\bf 26} (2009) 225002

\bibitem{anninos-anous}D. Anninos and T. Anous, \jhep {\bf 08} (2010) 131

\bibitem{zhang-zhao}J. Zhang and Z. Zhao, \prd {\bf 83} (2011) 064028

\bibitem{anninos-seng-strominger}D. Anninos, G. Seng Ng and A. Strominger, \cqg {\bf 28} (2011) 175019

\bibitem{kitamoto-kitazawa}H. Kitamoto and Y. Kitazawa, \npb {\bf 873} (2013) 325

\bibitem{mach-malec}P. Mach and E. Malec, \prd {\bf 88} (2013) 084055

\bibitem{german}G. German, A. Herrera-Aguilar, D. Malagon-Morejon, R. Mora-Luna and R. da Rocha, 

\jcap {\bf 02} (2013) 035

\bibitem{kodama-arraut}H. Kodama and I. Arraut, \ptep (2014) {023E02}

\bibitem{zhang-jun-wang}C-Y. Zhang, S-J. Zhang and B. Wang, arXiv:1405.3811 (2014)

\bibitem{paranjape}S. Mbarek and M. B. Paranjape, arXiv:1407.1457 [gr-qc] (2014)

\bibitem{modesto}L. Modesto, \prd {\bf 70} (2004) 124009

\bibitem{garfinkle-HS}D. Garfinkle, G.T. Horowitz and A. Strominger, \prd {\bf 43} (1991) 3140

\bibitem{kar-maharana-panda}S. Kar, J. Maharana and S. Panda, \npb {\bf 465} (1996) 439

\bibitem{polchinski}J. Polchinski, \prl {\bf 75} (1995) 4724

\bibitem{seiberg-witten}N. Seiberg and E. Witten, \jhep {\bf 09} (1999) 032

\bibitem{gibbons}G.W. Gibbons and K. Hashimoto, \jhep {\bf 09} (2000) 013; G.W. Gibbons and A. Ishibashi, \cqg {\bf 21} (2004) 2919

\bibitem{bachas}C. Bachas and M. Petropoulos, \jhep {\bf 0102} (2001) 025

\bibitem{mars}M. Mars, J.M.M. Senovilla and R. Vera, \prl {\bf 86} (2001) 4219

\bibitem{kar-panda}S. Kar and S. Panda, \jhep {\bf 0211} (2002) 052

\bibitem{kar-majumdar}S. Kar and S. Majumdar, \prd {\bf 74} (2006) 066003; \ijmpa {\bf 21} (2006) 6087; \ijmpa {\bf 21} (2006) 2391

\bibitem{kar-ds}S. Kar, \prd{\bf 74} (2006) 126002; \jhep {\bf 0610} (2006) 052; \ijmpa{\bf 24} (2009) 3571

\bibitem{liu}L-H. Liu, B. Wang and G-H. Yang, \prd {\bf 76} (2007) 064014

\bibitem{zhang2}J. Zhang, \plb{\bf 668} (2008) 353

\bibitem{grezia-esposito-miele}E.D. Grezia, G. Esposito and G. Miele, J.Phys.{\bf A} {\bf 41} (2008) 164063

\bibitem{kklt}S. Kachru, R. Kallosh, A. Linde and S. Trivedi, \prd {\bf 68} (2003) 046005

\bibitem{burgess-kallosh-quevedo}C. P. Burgess, R. Kallosh and F. Quevedo, \jhep {\bf 10} (2003) 056

\bibitem{keshav-tatar}K. Dasgupta, R. Gwyn, E. McDonough, M. Mia and R. Tatar, \jhep {\bf 1407} (2014) 054

\bibitem{hawking}S.W. Hawking, \cmp \ {}{\bf 43} (1975) 199

\bibitem{hawking-gibbons}G.W. Gibbons and S.W. Hawking, \prd {\bf 15} (1977) 2738

\bibitem{mottola}E. Mottola, \prd {\bf 31} (1985) 754
 
\bibitem{burgess-majumdar}C. P. Burgess, M. Majumdar, D. Nolte, F. Quenvedo, G. Rajesh and R-J. Zhang,

\jhep {\bf 07} (2001) 047

\bibitem{majumdar-davis}M. Majumdar and A-C. Davis, \jhep {\bf 03} (2002) 056

\bibitem{spsk-JHEP}A.K. Singh, K.P. Pandey, S. Singh and S. Kar, \jhep {\bf 05} (2013) 033

\bibitem{spsk-PRD}A.K. Singh, K.P. Pandey, S. Singh and S. Kar, \prd {\bf 88} (2013) 066001

\bibitem{spsk-NPB-P}A.K. Singh, K.P. Pandey, S. Singh and S. Kar, \npb {\bf 251-252} (2014) 241 

\bibitem{spsk-NPB1}S. Singh, K.P. Pandey, A.K. Singh and S. Kar, \npb {\bf 879} (2014) 216

\bibitem{spsk-NPB2}S. Singh, K.P. Pandey, A.K. Singh and S. Kar, \ijmpa (2014) in press, arXiv:1311.3604

\bibitem{pssk-jaat}K.P. Pandey, A.K. Singh, S. Singh and S. Kar, \jaat {\bf 03} (2013) 01; 

S. Kar, \jaat {\bf 03} (2013) e106

\bibitem{psskk}K.P. Pandey, A.K. Singh, S. Singh, R. Kapoor and S. Kar, arXiv:1405.3931 [hep-th] (2014)

\bibitem{pssk1}K.P. Pandey, A.K. Singh, S. Singh and S. Kar, arXiv:1405.6113 [hep-th] (2014)

\bibitem{pssk2}K.P. Pandey, A.K. Singh, S. Singh and S. Kar, arXiv:1405.7917 [hep-th] (2014)

\bibitem{kpss}S. Kar, K.P. Pandey, S. Singh and A.K. Singh, World Sci.Proceeding C10-02-24 (2011) 559; 

A.K. Singh, K.P. Pandey, S. Singh and S. Kar, World Sci.Proceeding C10-02-24 (2011) 567

\bibitem{thooft}G.'t Hooft, \npb {\bf 304} (1988) 867

\bibitem{kar-maharana}S. Kar and J. Maharana, \ijmpa {\bf 10} (1995) 2733

\bibitem{hanada}M. Hanada, arXiv:1407.5322 [hep-ph] (2014)

\bibitem{candelas-HS}S. Candelas, G.T. Horowitz and A. Strominger, \npb {\bf 258} (1985) 46

\bibitem{freed}D.S. Freed, \cmp {\bf 107} (1986) 483

\end{thebibliography}
\end{document}